\documentclass[prd,twocolumn,a4paper,superscriptaddress,floatfix]{revtex4}
\usepackage{graphicx}
\begin{document}

\newcommand{\be}{\begin{equation}}
\newcommand{\ee}{\end{equation}}
\newcommand{\bq}{\begin{eqnarray}}
\newcommand{\eq}{\end{eqnarray}}
\newcommand{\bsq}{\begin{subequations}}
\newcommand{\esq}{\end{subequations}}
\newcommand{\bc}{\begin{center}}
\newcommand{\ec}{\end{center}}
\newcommand {\R}{{\mathcal R}}
\newcommand{\al}{\alpha}
\newcommand\lsim{\mathrel{\rlap{\lower4pt\hbox{\hskip1pt$\sim$}} \raise1pt\hbox{$<$}}}
\newcommand\gsim{\mathrel{\rlap{\lower4pt\hbox{\hskip1pt$\sim$}} \raise1pt\hbox{$>$}}}

\title{Dynamics of domain wall networks with junctions}

\author{P.P. Avelino} 
\email[Electronic address: ]{ppavelin@fc.up.pt} 
\affiliation{Centro de F\'{\i}sica do Porto, Rua do Campo Alegre 687, 4169-007 Porto, Portugal} 
\affiliation{Departamento de F\'{\i}sica da Faculdade de Ci\^encias da Universidade do Porto, Rua do Campo Alegre 687, 4169-007 Porto, Portugal} 
\author{C.J.A.P. Martins} 
\email[Electronic address: ]{Carlos.Martins@astro.up.pt}
\affiliation{Centro de Astrof\'{\i}sica da Universidade do Porto, Rua das Estrelas s/n, 4150-762 Porto, Portugal}
\affiliation{DAMTP, University of Cambridge, Wilberforce Road, Cambridge CB3 0WA, United Kingdom} 
\author{J. Menezes}  
\email{jmenezes@fc.up.pt} 
\affiliation{Centro de F\'{\i}sica do Porto, Rua do Campo Alegre 687, 4169-007 Porto, Portugal} 
\affiliation{Departamento de F\'\i sica, Universidade Federal da Para\'\i ba, Caixa Postal 5008, 58051-970 Jo\~ao Pessoa, Para\'\i ba, Brazil} 
\author{R. Menezes} 
\email{rms@fisica.ufpb.br} 
\affiliation{Departamento de F\'\i sica, Universidade Federal da Para\'\i ba, Caixa Postal 5008, 58051-970 Jo\~ao Pessoa, Para\'\i ba, Brazil} 
\author{J.C.R.E. Oliveira} 
\email{jeolivei@fc.up.pt} 
\affiliation{Centro de F\'{\i}sica do Porto, Rua do Campo Alegre 687, 4169-007 Porto, Portugal} 
\affiliation{Departamento de F\'{\i}sica da Faculdade de Ci\^encias da Universidade do Porto, Rua do Campo Alegre 687, 4169-007 Porto, Portugal} 

\date{24 July 2008}

\begin{abstract}
We use a combination of analytic tools and an extensive set of the largest and most accurate three-dimensional field theory numerical simulations to study the dynamics of domain wall networks with junctions. We build upon our previous work and consider a class of models which, in the limit of large number $N$ of coupled scalar fields, approaches the so-called `ideal' model (in terms of its potential to lead to network frustration). We consider values of $N$ between $N=2$ and $N=20$, and a range of cosmological epochs, and we also compare this class of models with other toy models used in the past. In all cases we find compelling evidence for a gradual approach to scaling, strongly supporting our no-frustration conjecture. We also discuss the various possible types of junctions (including cases where there is a hierarchy of them) and their roles in the dynamics of the network. Finally, we revise the Zel'dovich bound and provide an updated cosmological bound on the energy scale of this type of defect network: it must be lower than $10 \, {\rm keV}$.
\end{abstract}
\maketitle

\section{\label{sint}Introduction}

As the early universe expanded and cooled down, it is believed to have gone through a series of phase transitions, at which networks of topological defects must necessarily have formed \cite{KIBBLE}. The type of defect that forms and its specific properties depend on the particular details of each symmetry breaking, and hence there is a wide range of possibilities, which will lead to correspondingly very different cosmological consequences \cite{VSH}. Domain walls are known to be pathological except if they are very light \cite{ZEL}, but it has been claimed \cite{SOLID} that if a domain wall network is frozen in co-moving coordinates (or `frustrates', as is often colloquially put) then it can naturally explain the observational evidence that points to a recent acceleration of the universe. 

It is clear that this scenario is subject to a number of observational constraints. For example the cosmic microwave background (CMB) data \cite{WMAP5} severely constrains the characteristic scale of the network, $L$, which needs to be tiny in order not to give rise to exceedingly large CMB fluctuations. Also, recalling that the equation of state of a domain wall gas is given by
\begin{equation}
w \equiv \frac{p}{\rho} = -\frac{2}{3} + v^2\,,
\end{equation}
with $v$ being the root-mean squared (RMS) velocity of the walls and that we require that
\begin{equation}
w < -\frac{1}{3}\left(1+\frac{\Omega_m^0}{\Omega_{DE}^0}\right) \lsim -\frac{1}{2}
\end{equation}
in order to accelerate the universe at the present time it is clear that only a non-relativistic value of the velocity (in fact, basically $v\sim0$) would have any chance of working. The simplest domain wall models are known to reach a scaling regime (until they dominate the energy density of the universe), as first pointed out in \cite{PRESS} and recently studied in detail in \cite{AWALL,SIMS1,SIMS2}, and hence are obviously unable to satisfy these constraints. Nevertheless, it was thought that more complicated models, notably those having junctions, would eventually frustrate and therefore might conceivably be able to satisfy them.

In previous work \cite{IDEAL1,IDEAL2} we have studied the dynamics of domain wall networks with junctions, and investigated in detail energy, geometrical and topological constraints on their properties. This led us to develop an ideal class of models which includes, in the large $N$ limit, what we called the \textit{ideal model} (that is, the best possible candidate for frustration). A series of analytic and numerical arguments then led us to a \textit{no frustration conjecture}: even though one can build (purely by hand, as was done in \cite{BATTYE,CARTER,IDEAL2}) special lattices that would be locally stable against small perturbations, no such configurations are expected to ever emerge from any realistic cosmological phase transition. (A much simpler analysis of the evolution of non-interacting and entangled cosmic string networks using a velocity-dependent one-scale model had already led to a similar result, at least in four space-time dimensions \cite{NONINT}.) Our high-resolution numerical simulations of various realizations of the ideal class and other models showed clear evidence of a gradual approach to scaling, which was subsequently confirmed in \cite{BMFLAT} (though note that the latter only reports on numerical simulations in Minkowski space thus neglecting the role of the expansion). 
 
Still, a fair criticism of our earlier work \cite{IDEAL1,IDEAL2} would be that we only considered numerical simulations in two spatial dimensions, and the behavior could well be different in the three spatial dimensions of our universe. More recently \cite{THIRD} we endeavored to eliminate this shortcoming and presented results of a series of massively parallel domain wall network numerical simulations of the `ideal' class of models in three spatial dimensions, which confirmed our earlier work, providing conclusive evidence for a gradual approach to scaling and hence strongly supporting our no frustration conjecture. The present article is the fourth in this series of papers, and its aims are to present the results of \cite{THIRD} more thoroughly, to extend them in a number of different ways, and to discuss in some detail the corresponding cosmological implications.

In Sect. \ref{swal} we introduce various approaches to the study of domain walls---the microscopic description, the analytic modeling and the building of phenomenological models---that we will be using in the rest of the paper. In particular we highlight some relevant scaling solutions for domain wall networks and describe the ideal class of models which will be our primary focus. In Sect. \ref{sfield} we discuss several technical details of our massively parallel field theory numerical code (which uses an improved version of the algorithm of Press, Ryden and Spergel \cite{PRESS}, henceforth referred to as the PRS algorithm). In particular we describe in some detail how we identify the domain walls and how we measure their velocities. Sect. \ref{ssim} contains our main numerical results: we describe the outcome of our various sets of three-dimensional high-resolution simulations of the ideal class of models. We also discuss the relevance of the key findings and contrast the results with those of other models. This then leads us to discuss, in Sect. \ref{smod}, the analytic modeling of our results, including the cosmologically important asymptotic limit of the ideal class of models. In Sect. \ref{sjun} we present an analysis of the possible hierarchies of junctions in the ideal and other models, as well as a brief discussion of the role of the junctions on the dynamics of the networks. Finally Sect. \ref{scon} has our conclusions, as well as some comments on the implications of our results for cosmological scenarios involving domain walls and a brief outline of future endeavors in this area.


\section{\label{swal}Modeling domain walls}

We shall start by describing some relevant micro-physical and averaged (macro-physical) tools for modeling domain walls and their evolution. While parts of this section are a review of previous work (see for example \cite{VSH}), they will be useful not only as a means to set up some notation but also as a basis upon which we will build the discussion in subsequent sections.

\subsection{\label{basicdef}What is a domain wall?}

Let us start by a basic but important point: defining what a cosmological domain wall is in a simple but relatively generic and physically meaningful way. A cosmological domain wall is effectively a two-dimensional object, meaning that its thickness is much smaller than the curvature radius. This implies that locally the domain wall is planar. Consider a local inertial reference frame in which the domain wall is instantaneously at rest and assume that the domain wall is perpendicular to the $z$ direction. Another defining characteristic of a cosmological domain wall is that its properties are independent of $x$ and $y$, which means that both its thickness, $\delta$, and its mass per unit area, $\sigma$, do not change along the wall. The fact that nothing changes along the wall implies that the physical velocity of the domain wall is perpendicular to it (it is not possible to measure parallel velocities).

These simple properties are sufficient to determine their energy-momentum tensor \cite{KOLB}. Let us start by considering a domain wall network which is frozen in co-moving coordinates in a Friedmann-Robertson-Walker (FRW) universe, and let us assume without loss of generality that the characteristic length is small enough that the network can be considered homogeneous and isotropic on cosmological scales. If that is the case then the average domain wall density is
\begin{equation}
\rho = \frac{\sigma A}{V} \propto \frac{a^2}{a^3} \propto a^{-1}\, ,
\end{equation}
where $V$ is a large volume and $A$ is the total domain wall area inside $V$.
Using the fact that
\begin{equation}
\frac{d \rho}{dt} + (1+w)H \rho=0 \, ,
\end{equation}
we get that $w=-2/3$ for a frozen domain wall network. 

Let us now move from the domain wall network as a whole and try to compute the energy momentum tensor $T^{\alpha \beta}$ of the individual domain wall segments. Consider a local inertial frame in which the domain wall is instantaneously at rest and assume that the wall is perpendicular to the $z$ direction. In this case, we can show that
\begin{equation}
T_{\mu\nu}=diag\left[\rho_w(z),-\rho_w(z),-\rho_w(z),0\right]\label{eq:tensorparedeplana}
\end{equation}
where $\rho_w$ is the wall energy-density and $T_{ii}$ are the stress components along the $i$ direction. The fact that the wall properties do not change along it means that it must be impossible to measure velocities in any direction parallel to the wall plane. So $T_{\mu\nu}$ must be the invariant with respect to Lorentz boosts in any direction along the wall.

Consider the following boost in the $x$ direction
\begin{equation}
t'=\gamma(t+vx)\,,\quad x'=\gamma(x+vt)\,,\quad y'=y\,,\quad z'=z
\end{equation}
where $\gamma=\left(1-v^{2}\right)^{-1/2}$ is the Lorentz factor and $v$ is the wall velocity. Hence $T_{\mu\nu}$ transforms
as
\begin{equation}
T^{\mu'\nu'}=\Lambda_{\alpha}^{\mu'}\Lambda_{\beta}^{\nu'}T^{\alpha\beta}\label{eq:transfT}
\end{equation}
where
\begin{equation}
\Lambda_{t}^{t'}=\Lambda_{x}^{x'}=\gamma\,,\quad\Lambda_{x}^{t'}=\Lambda_{t}^{x'}=\gamma v\,,\quad\Lambda_{y}^{y'}=\Lambda_{z}^{z'}=1
\end{equation}
and all other components vanish. By requiring that
\begin{equation}
T^{t'y'}=\gamma T^{ty}+\gamma vT^{xy}=T^{ty}
\end{equation}
and
\begin{equation}
T^{x'y'}=\gamma T^{xy}+\gamma vT^{ty}=T^{xy}
\end{equation}
we obtain $T^{ty}=T^{xy}=0$. Similarly from
\begin{equation}
T^{t'z'}=\gamma T^{tz}+\gamma vT^{xz}=T^{tz}
\end{equation}
and
\begin{equation}
T^{x'z'}=\gamma T^{xz}+\gamma vT^{ty}=T^{xz}
\end{equation}
we have $T^{tz}=T^{xz}=0$. On the other hand, 
\begin{equation}
T^{t't'}=\gamma T^{tt}+\gamma^{2}v^{2}T^{xx}=T^{tt}
\end{equation}
implies that $T^{xx}=-T^{tt}$. From the fact that there is no preferred direction along the wall we have $T^{xt}=T^{yt}$, $T^{xz}=T^{yz}$, and $T^{xx}=T^{yy}$. Finally from the energy-momentum conservation law we have, $T^{\mu\alpha}{}_{,\alpha}=0$, and so $T^{zz}{}_{,z}=0$. This is equivalent to $T^{zz}=0$ since $T^{zz}$ is null, at $z\rightarrow\infty$, for any domain wall solution. We then conclude that a planar domain wall solution has the energy-momentum tensor given by Eq. \ref{eq:tensorparedeplana}. Consequently, given $T^{00}$ all other components of the energy-momentum tensor of the domain wall are uniquely determined. Note that all of this is independent of the microscopic structure of the domain wall.

This has an important consequences for domain wall dynamics. Given the above considerations, a domain wall is essentially characterized by its mass per unit area $\sigma$ (which is fixed) and its velocity. Therefore there is only one variable to evolve, $v$, and its evolution is fully determined by energy-momentum conservation ${T^{\alpha \beta}}_{; \beta}=0$. In the particular case of a planar domain wall one has
\begin{equation}
\gamma v \propto a^{-3}\,.
\end{equation}

Now consider the role of friction effects on the dynamics of domain walls. Since we are mainly interested in showing that no realistic domain wall network will frustrate by the present epoch we need only consider the maximum frictional force consistent with energy-momentum conservation. For interaction with radiation one has
\begin{equation}
\left(\frac{dv}{dt}\right)_{max} \sim -\frac{\rho_{rad}}{\sigma} v \, ,
\end{equation}
while for matter we have
\begin{equation}
\left(\frac{dv}{dt}\right)_{max} \sim -\frac{\rho_{mat}}{\sigma} v^2 \, ,
\end{equation}
and in both of these we are neglecting relativistic corrections. The extra factor of $v$ which appears in the interactions with matter gives us an extra reason not consider them in the analysis. In our previous work \cite{IDEAL1,IDEAL2} we justified not considering interactions with matter on the grounds that they would adversely affect structure formation. However, there could still be some fraction of the dark matter in the universe (say one part in a hundred) with which domain walls could interact without causing obvious observable effects. However, that interaction in the small $v$ limit (the one we are interested in) would be much weaker than in the radiation case. This justifies, and indeed strengthens our conclusions.

\subsection{\label{Lagran}Generalized Lagrangians}

In the above we considered some properties of domain walls which are independent of the particular scalar field model we were considering. But are the above conclusions regarding domain wall dynamics valid more generally? It turns out that a more general argument can indeed be given.

Consider a single-field model 
\be 
{\cal S}=\int dt \int d^3 x \,\sqrt{-g} \,{\cal L}(\phi, X) 
\ee  
where $X=1/2 \partial_\mu \phi \partial^\mu \phi$. Varying the action with respect to $\phi$ we get the following equation of motion 
\be 
\frac{1}{\sqrt{-g}} \,\partial_\mu \left(\sqrt{-g} \,{\cal L}_X \partial ^\mu \phi \right) = {\cal L}_\phi 
\ee  
where ${\cal L}_X=\frac{\partial {\cal L}}{\partial X}$ and ${\cal L}_\phi=\frac{\partial {\cal L}}{\partial \phi}$. The FRW metric has been assumed. Re-writing this in physical coordinates we have 
\be 
\frac{\partial}{\partial t}\left({\cal L}_X  \frac{\partial \phi}{\partial t}\right) + 3H{\cal L}_X  \frac{\partial \phi}{\partial t} - \frac{\partial}{\partial x_i} \left({\cal L}_X \frac{\partial \phi}{\partial x_i}\right) = {\cal L}_\phi 
\ee  
and for planar domain wall solutions (depending only on one spatial coordinate, $z$) this can be re-written 
\be\label{eq1234} 
\frac{\partial}{\partial t}\left({\cal L}_X  \frac{\partial \phi}{\partial t}\right) + 3H{\cal L}_X  \frac{\partial \phi}{\partial t} - \frac{\partial}{\partial z} \left({\cal L}_X \frac{\partial \phi}{\partial z}\right) = {\cal L}_\phi\,, 
\ee  
which has a solution $\phi(\theta)$ with $\theta=z$.
For a static wall, this reduces to  
\be 
\label{staticwall} 
- \frac{\partial}{\partial z} \left({\cal L}_X \frac{\partial \phi}{\partial z}\right) = {\cal L}_\phi\,. 
\ee  
Now consider a wall moving with speed $v$ in the $z$ direction. Then
\be 
\frac{\partial }{\partial t}=\frac{\partial \theta}{\partial t}\frac{d}{d\theta}  
=v \gamma\,\frac{d}{d\theta} \,,
\ee
\be\frac{\partial }{\partial z}=\frac{\partial \theta}{\partial z}\frac{d}{d\theta}  
=\gamma\frac{d}{d\theta} \,; 
\ee 
noting that in this case $X=-\frac12 (\frac{d\phi}{d\theta})^2$, one can write 
\be 
- \frac{d}{d\theta} \left({\cal L}_X \frac{d \phi}{d \theta}\right)+\frac{d\phi}{d\theta} {\cal L}_X \left[ \frac{d(v\gamma)}{dt}+3H(v\gamma)\right]= {\cal L}_\phi 
\ee  
and given that $\phi(\theta)$ is a solution to Eqn. \ref{staticwall}, we must have 
\be\label{relation} 
\frac{d(v\gamma)}{dt}+3H(v\gamma)=0 
\ee 
which again leads us to $\gamma v \propto a^{-3}$, generalizing the result in the previous sub-section. 
 
In models with multiple scalar fields there are other ways of putting together derivative terms which preserve Lorentz invariance, for example $\partial_\mu \phi \partial^\mu \chi$, but a similar analysis would still lead to Eqn. \ref{relation}. There are, however, ways to circumvent this constraint. The most obvious is to relax the FRW metric assumption, or in other words moving away from standard cosmology. One example would be to consider models with extra dimensions.

\subsection{\label{vosmodel}Simple analytic modeling}

In \cite{AWALL} we introduced a phenomenological one-scale model for the evolution of domain wall networks that has been shown to provide a good approximation to the evolution of two key network parameters: the characteristic scale of the network, $L$, and the RMS velocity of the domain walls, $v$. Their evolution equations are
\begin{equation}
\frac{dL}{dt}=HL+\frac{L}{\ell_d}v^2+cv\,,
\label{rhoevoldw1}
\end{equation}
\begin{equation}
\frac{dv}{dt}=(1-v^2)\left(\frac{k}{L}-\frac{v}{\ell_d}\right)\,,
\label{rhoevoldw2}
\end{equation}
where $H$ is the Hubble parameter, $c$ is the energy loss efficiency, $k$ is the curvature parameter and we have defined a damping length scale,
\begin{equation}
\frac{1}{\ell_d}=3H+\frac{1}{\ell_f}\,,
\end{equation}
which includes both the effects of Hubble damping and particle scattering. The characteristic scale of the network is defined as $L=\sigma / \rho$ where $\rho$ is the average density in domain walls and $\sigma$ is the wall mass per unit area. Note that if domain walls are an important contribution to the dark energy then $\rho$ must be of the order of the critical density, $\rho_c$, at the present time.

Just as in the case of cosmic strings, the generic attractor of the above equations in a cosmological model where the scale factor grows as $a\propto t^\alpha$ is a linear scaling solution (although in the case of the domain walls they will eventually dominate the energy density of the universe), that is characterized by the following (constant) parameters
\begin{equation}
\epsilon^2\equiv\left(\frac{L}{t}\right)^2=\frac{k(k+c)}{3\alpha (1-\alpha)}\, \label{defscalingg}
\end{equation}
\begin{equation}
v^2=\frac{1-\alpha}{3\alpha}\frac{k}{k+c}\,. \label{defscalingv}
\end{equation}
If we ignore the energy loss by the network (by making $c=0$) it is easy to show that a linear scaling solution is still possible for $\alpha > 1/4$. In the radiation era we obtain 
\begin{equation}
L=\frac{2}{\sqrt3}\, kt
\end{equation} 
\begin{equation}
v=\frac{1}{{\sqrt 3}}\,, \label{zerocrad}
\end{equation}
while in the matter era we have 
\begin{equation}
L=\sqrt {\frac{3}{2}}\, kt
\end{equation}
\begin{equation}
v=\frac{1}{{\sqrt 6}}\,. \label{zerocmat}
\end{equation}
Notice that in both eras we have $L \sim kt$ and relatively large velocities. If we require CMB temperature fluctuations generated by domain walls on scales of the order of Hubble radius to be smaller than $10^{-5}$ then one would need $(LH)^{3/2} \lsim  10^{-5}$ or equivalently $L \lsim 1 \, {\rm Mpc}$ (our conservative estimate in \cite{IDEAL1}). However, CMB observations imply that the fluctuations generated by the domain walls have to be smaller than $10^{-5}$ down to much smaller scales (say $\sim H^{-1}/100$). This means that current constraints on $L$ are expected to be roughly $2$ orders of magnitude stronger,
\begin{equation}
L \lsim 10 \, {\rm kpc}
\end{equation}
which implies a very small curvature parameter
\begin{equation}
k \lsim 10^{-6}\,.
\end{equation}
This clearly shows that the simplest domain wall scenario without junctions (which necessarily has $k \sim 1$) is ruled out as a dark energy scenario, regardless of any other considerations. It is easy to show that allowing for a non-zero $c$ leads to a larger $L$ and consequently it does not help frustration \cite{AWALL,IDEAL1}. On the other hand, as was pointed out above, including friction also does not help much, due to the limited amount of energy with which domain walls can interact conserving energy and momentum \cite{IDEAL1}.

\subsection{\label{idealmodel}Towards the ideal model}

This failure of the simplest domain wall scenario led us to consider more complex scenarios with junctions and in \cite{IDEAL1} we set out to investigate in detail energy, geometrical and topological considerations that severely constrain the properties of domain wall networks. In particular, in the context of 2D domain wall networks, it was shown using local energy considerations that two-edge domains are always unstable and that three-, four- and five-edge domains will be unstable if only Y-type junctions occur in a given model. We have also demonstrated that increasing the average dimensionality of the junctions, $\langle d \rangle$, leads to a decrease of the average number of edges, $\langle x \rangle$, per domain (in particular if $\langle d \rangle > 6$ then $\langle x \rangle < 3$ and consequently no equilibrium configurations will ever form). Moreover, allowing for domain walls with different tensions contributes to increasing the instability (relative to models where all the walls have the same tension). This is because the walls with higher tension will tend to collapse, thereby increasing the dimensionality of the junctions which, in turn, will in general lead to the production of further unstable two edge domains.

Another important aspect is related to the fact that the stability of a given domain depends on global considerations (those depending on the configuration of the surrounding domains) as well as local ones (those associated with the domain itself) and we expect the global ones to become more important as we increase the dimensionality of the junctions or consider specific domains with a large number of edges. In particular it is possible to show (in 2D) that a domain with three edges only survives the local stability analysis by very little in the case where there are only $X$ type junctions (the potential energy after the collapse would be at most about $10\%$ larger than before). In this case, we expect that, in general, non-local effects will make a three edge domain unstable. As a result in a model with only $X$ type junctions the only possible stable configuration is the one in which all the domains have the same number of edges (namely four), which never occurs in the context of realistic domain wall network simulations. Note that we are assuming the junctions to be free throughout the paper. Otherwise their energy-momentum contribution could not be neglected, spoiling the dark energy properties associated with a static domain wall network. We shall return to the role of the junctions later in this paper.

By combining energy, geometrical and topological considerations, one is then naturally led to a class of models with any number $N$ scalar fields and correspondingly $N+1$ vacua, with the property that all possible domain walls have equal tensions \cite{IDEAL2}. The  `ideal' model, meaning the one with maximal \textit{a priori} probability of reaching a frustrated state, is then obtained in the limit $N \to \infty$. For large $N$ the collapse of a single domain will only very rarely lead to the fusion of two of the surrounding domains. This is clearly a very desirable feature from the point of view of frustration (see \cite{IDEAL1}). Also, by requiring all domain walls to have equal tensions we avoid another potential source of instability.

A specific realization of this class of models with $N$ scalar fields and a scalar field potential with $N+1$ vacua was first given in \cite{IDEAL2}
\begin{equation}
V \propto \sum_{j=1}^{N+1} r_j^2 \left(r_j^2 - r_0^2\right)^2\ {\rm with}\ 
r_j^2=\sum_{i=1}^N (\phi_i - p_{{i}_{j}})^2\,, \label{ideal}
\end{equation}
where $p_{{i}_{j}}$ are the $N+1$ coordinates of the vacua of the potential. We have chosen $p_{{i}_{j}}$ to be the vertices of an ($N+1$)-dimensional regular polyhedron, and fixed the distance between the vacua to be equal to the parameter $r_0$. Given that in this model all possible domain walls have equal tensions, only Y-type junctions will form. Although it is possible to have only $X$ type junctions in a model with $4$ minima, this is in general not the case for a larger number of minima if all possible vacuum configurations are allowed to exist. In particular this means that it is not possible to construct a generalization of the `ideal' model in which all junctions are of the $X$ type. 

In \cite{IDEAL2} we performed 2D field theory simulations of the `ideal' class of models with $N=4$ and $N=7$. These results are useful since they are much simpler (hence easier to understand), containing many important features which are also relevant in higher dimensions. However, realistic 3D domain wall network simulations were required in order to test our \textit{no frustration conjecture}. The results of a set of 3D matter era simulations was first presented in \cite{THIRD}, confirming the above expectations. In the following sections we will describe the algorithm used to obtain these in somewhat more detail, and present and discuss the results of a much more extensive set of simulations,


\section{\label{sfield}Massively parallel simulations}

We have performed high-resolution field theory numerical simulations of domain wall networks on the UK Computational Cosmology Consortium's COSMOS supercomputer using a modified version of the PRS \cite{PRESS}. Earlier descriptions of our code can be found in \cite{SIMS1,SIMS2}, and some more specific details are described below. The numerical code, in C language, was parallelized with OpenMP directives, and gradually optimized for the shared memory architecture of COSMOS. The code plus its auxiliary routines are about 5000 lines long.

The results to be discussed below come mainly from simulations of $128^3$, $256^3$ or $512^3$ boxes, in various cosmological epochs (matter domination, radiation domination and so on) for the ideal class of models with all values of $N$ between 2 and 20. In addition, some realizations of other models were also performed (these particular models will be described as they are introduced). Unless otherwise is stated, we assume initial conditions where the scalar field at each point in the grid is associated with a randomly chosen minimum of the potential.

The required memory for our three-dimensional simulations is approximately
\begin{equation}
MEM=N*4.5*8*\left(\frac{DIM}{1024}\right)^3 GB\,;
\end{equation}
as an example, for $512^3$ boxes 90 GB are required for 20 fields. An output box binary file can also be produced at specified time steps which can then be used to generate animations: an example of these is available at \url{http://www.damtp.cam.ac.uk/cosmos/viz/movies/evo2_25620_msmpeg.avi} (several others are available from the authors upon request). As a benchmarking example, a $512^3$ simulation with 3 fields (requiring about 14.5 Gb of memory) takes about 14 seconds per step on 16 processors and only 5 seconds per step on 32 processors (as the memory ratio becomes favorable) without box output, and a complete run takes just 85 minutes. For larger runs the scalability is good if one keeps the memory smaller than 1 GB per processor. The example given above, a $512^3$ box with 20 scalar fields and box outputs at every time step, takes about 2 hours and 15 minutes on 128 processors.

\subsection{\label{measurearea}Identifying the walls}

We recall that the key idea behind the PRS algorithm \cite{PRESS} is to modify the domain wall thickness in order to ensure a fixed comoving resolution, thus enabling much larger dynamic ranges to be simulated. We defined the domain wall as the region where
\begin{equation}
V(\phi) > \alpha V_{\rm max}\,, \label{defnalpha}
\end{equation}
$V_{\rm max}$ being the maximum of the potential and $0 < \alpha < 1$. The chosen threshold value $\alpha$ will thus define two properties of the network: the corresponding thickness of the static domain wall $\delta$ and the volume fraction of the box with domain walls $f$. They can be used to estimate the \textit{comoving characteristic scale} of the network
\begin{equation}
L_c \equiv \frac{L}{a} \sim\frac{\delta}{f}
\end{equation}
where $f$ is the volume fraction of the box with domain walls (that is, satisfying the above condition), and $\delta$ is the thickness of a static domain wall. 

\begin{figure}
\includegraphics[width=3in]{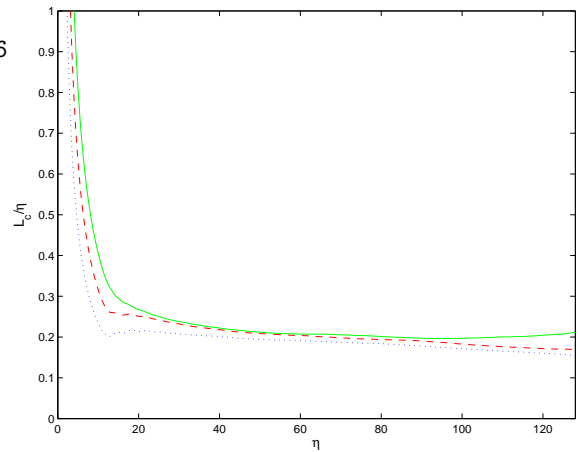}
\caption{\label{threshold}The relative comoving characteristic scale, $L_c/\eta $, for various values of the parameter $\alpha$ defined in Eqn. \protect\ref{defnalpha}. From bottom to top these are respectively $\alpha=0.6$ (blue, dotted line), $\alpha=0.4$ (red, dashed line) and $\alpha=0.2$ (green, solid line). All of these were obtained for $128^3$ simulations.}
\end{figure}
We have verified that for $0.2 \lsim \alpha \lsim 0.6$ (see Fig.\ref{threshold}), our results are almost independent of the threshold $\alpha$, as long as the domain wall is sufficiently resolved.  

In order to check whether the domain wall network is evolving according to a linear scaling solution we
define the \textit{scaling exponent} $\lambda$ so that
\be
L_c \propto \eta^{1-\lambda}\,,
\ee
where $\eta$ is the conformal time. If $\lambda = 0$ we have linear scaling, whereas the network frustrates for $\lambda=1$.
The scaling exponent can be calculated at two different values of the conformal time $\eta_1$ and $\eta_2$ by using the following relation \cite{SIMS1}
\be
\lambda(\eta_1,\eta_2) = \frac {\ln{(R_1/R_2)}} {\ln{(\eta_1/\eta_2)}},
\ee
where $R \equiv \eta\,L_c^{-1}$. In general, since the simulations are evolved until $\eta$ becomes equal to one half of the box comoving size (because beyond that point the effect of the finite box size will contaminate the results) the values of $\eta_1$ and $\eta_2$ are chosen in order to consider the second half of the dynamical range of the simulations. Note that the earlier part is not included to avoid contamination of the results by our particular choice of initial conditions, while beyond one half of the comoving size, the effect of the finite box size will contaminate the results.

\subsection{\label{measurev}Measuring wall velocities}

We measure the domain wall velocities using an algorithm analogous to that described in \cite{AWALL} which separates most of the radiated energy from the walls. We emphasize that this is mandatory for any reliable velocity measurement, since otherwise this radiation would contaminate the estimates of the velocities---our previous work shows that the contamination can easily be at the twenty percent level. This algorithm therefore represents an important advantage over previous velocity estimations (such as the one in the original PRS algorithm \cite{PRESS}). To be specific, we estimate the velocities as
\begin{equation}
v_*^2 \equiv \langle v^2 \gamma^2 \rangle = \sum_{V(\phi_i) > \alpha V_{\rm max}}  \frac{{\dot \phi_i}^2}{2 V(\phi_i)}\, ,
\end{equation}
where a dot represents a derivative with respect to conformal time and $\gamma=(1-v^2)^{1/2}$. 

Let us revisit our discussion in Sect. \ref{basicdef} showing that $\gamma v \propto a^{-3}$ for any planar domain wall perpendicular to the $zz$ direction and that our method for determining the RMS velocity of the domain walls will work independently of the particular scalar field potential we are considering. Let us write the solution for a moving wall as $\phi(\theta)$ where $\partial\theta/\partial t=\gamma v$ and $\partial\theta/\partial z=\gamma$. Then we have
\begin{equation}
\frac{\partial\phi}{\partial t}=\frac{d\phi}{d\theta}\gamma v\,\quad\frac{\partial\phi}{\partial z}=\frac{d\phi}{d\theta}\gamma\label{eq:transfsegundoz}
\end{equation}
and from the fact that $\phi(\theta)$ (with $\theta=z$) is a static solution we have
\begin{equation}
\frac{d^{2}\phi}{d\theta^{2}}=\frac{dV}{d\phi}\,;\label{eq:eqmov-estatica}
\end{equation}
we immediately obtain
\begin{equation}
\frac{d(v\gamma)}{dt}+3H(v\gamma)=0\,,
\end{equation}
which has the solution $\gamma v\propto a^{-3}$ as claimed. Note that by integrating
\begin{equation}
\frac{\partial^{2}\phi}{\partial\theta^{2}}=\frac{dV}{d\phi}\,,\end{equation}
we obtain
\begin{equation}
\left(\frac{\partial\phi}{\partial\theta}\right)^{2}=2V\,,
\end{equation}
and consequently it is straightforward to show that
\begin{equation}
v^{2}\gamma^{2}=\frac{1}{2V}\left(\frac{\partial\phi}{\partial t}\right)^{2}
\end{equation}
independently of the model. Applying the PRS procedure is equivalent to making the substitution $t\to\eta$, $z\to q_{z}$ in the above equations, where $dz=a(t)dq_{z}$ and $dt=a(t)d\eta$.


\section{\label{ssim}Numerical simulation results}

We are now in a position to present the bulk of our work: the most extensive, most accurate and largest (in terms of the conformal time dynamic range, which is the meaningful diagnostic) set of 3D field theory simulations of domain wall networks to date. We will start by describing the simulations of the ideal class of models in the matter era (which is the relevant context for discussing the possible role of domain wall networks as dark energy). Later in this section we also discuss simulations of the ideal class of models in other cosmological epochs, as well as some representative cases of simulations of other 'non-ideal' models: we will consider some examples from the phenomenological models of Kubotani \cite{KUBOTANI} and of Bazeia \textit{et al.} \cite{BAZEIA} (to which we will henceforth refer as the BBL model). When appropriate we will also compare these results with those obtained for simulations in other dimensions \cite{IDEAL1,IDEAL2} as well as those obtained for the simplest model of domain walls \cite{SIMS1,SIMS2,AWALL}, which have no junctions.

\subsection{\label{runsmat}Ideal model, matter era}

\begin{figure}
\includegraphics[width=3in]{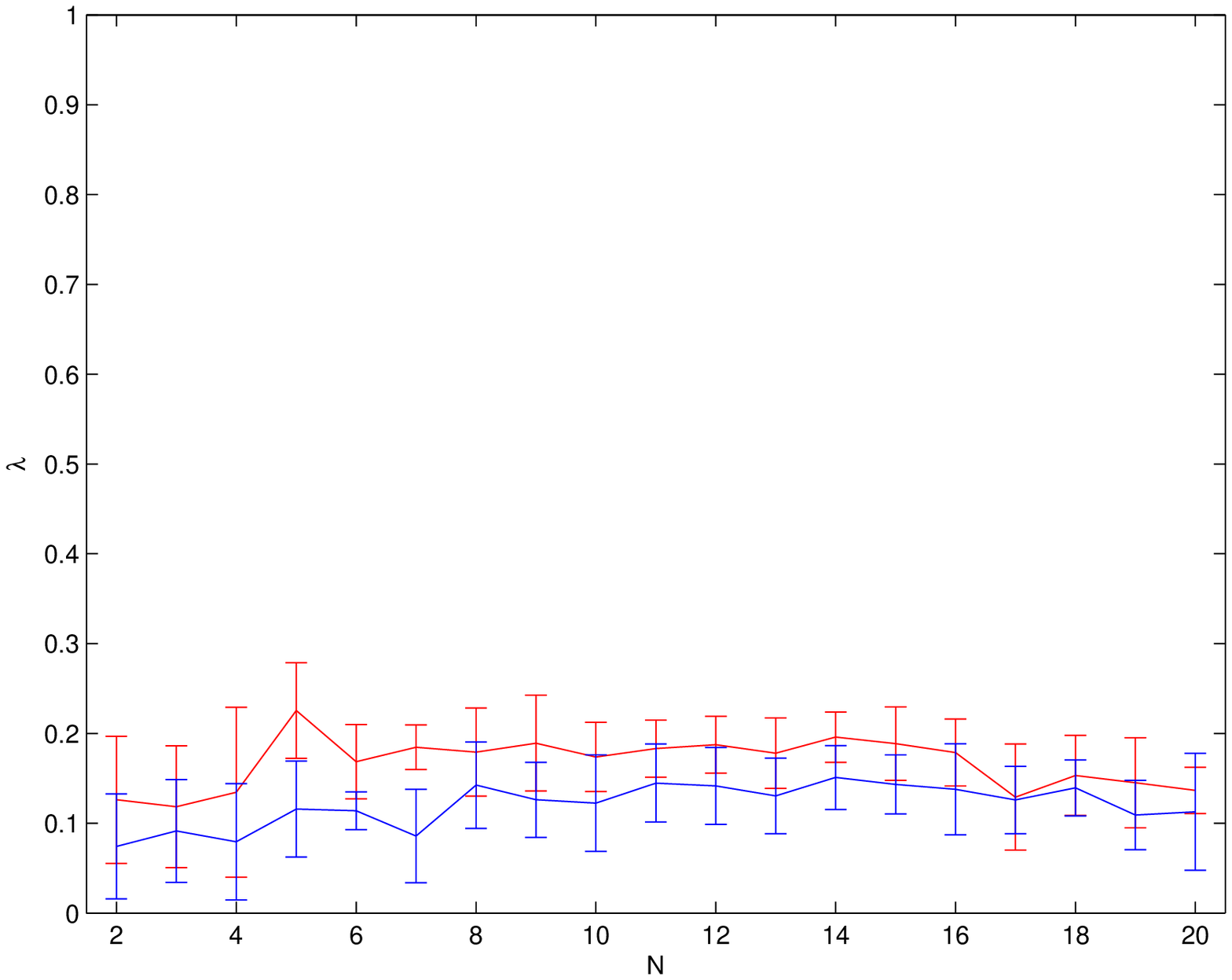}
\includegraphics[width=3in]{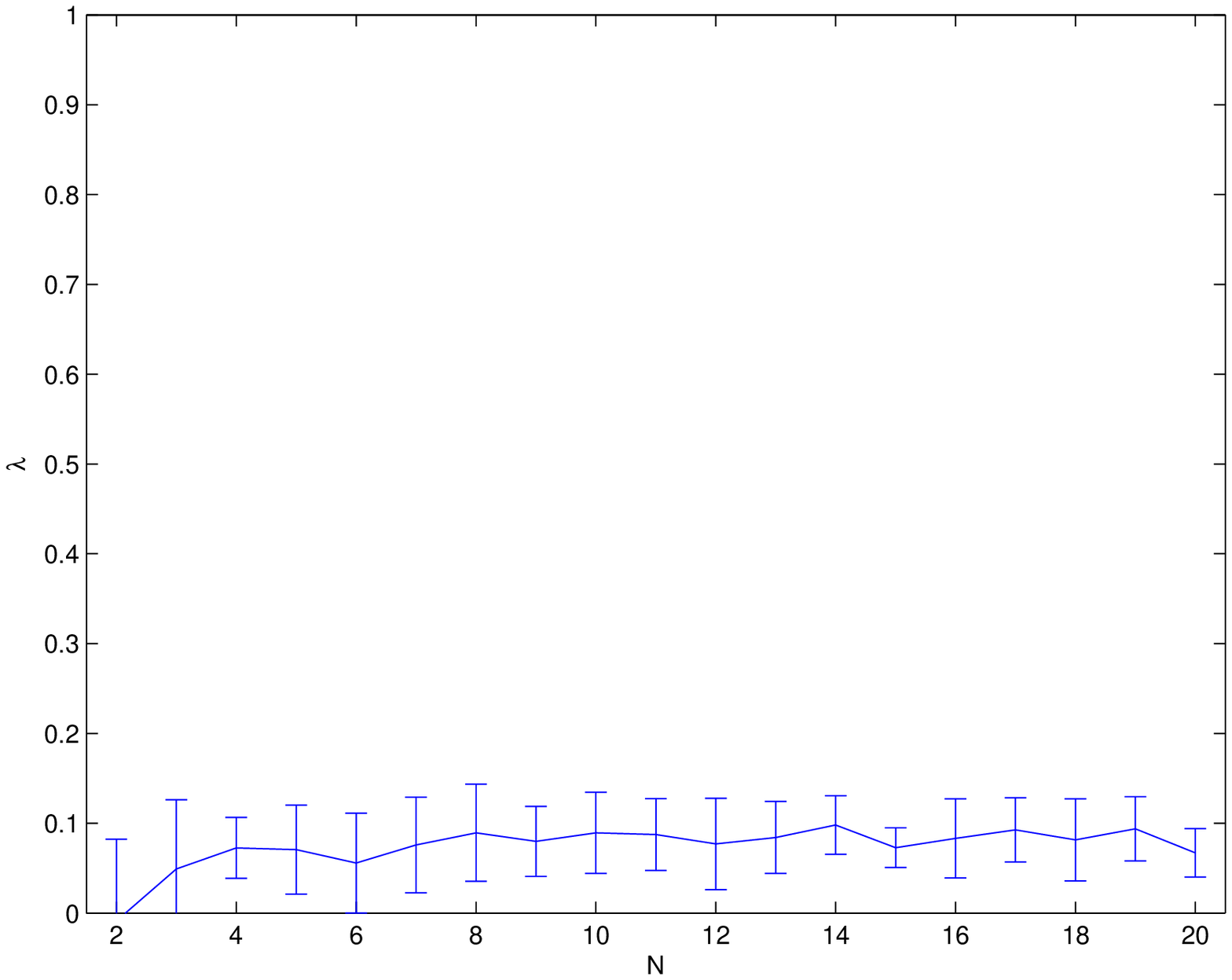}
\caption{\label{exponmat} The scaling exponents, $\lambda$, for all $N$'s between 2 and 20, for the $128^3$, $256^3$ boxes (top panel, top and bottom set of lines respectively) and $512^3$ boxes (bottom panel) in the matter-dominated era. The error bars represent the standard deviation in an ensemble of 10 simulations. Note that frustration would correspond to $\lambda=1$.}
\end{figure}

In Fig. \ref{exponmat} we plot the scaling exponents, $\lambda$, defined by
\begin{equation}
\frac{L_c}{\eta} \propto \eta^{-\lambda}\,,\label{comscaldef}
\end{equation}
for the ideal class of models with all numbers of fields $N$ between 2 and 20, in the matter-dominated era. We show results of simulations in $128^3$, $256^3$ and $512^3$ boxes, to illustrate the effect of the box size (and hence dynamic range) on the results---these effects are small but still identifiable, even with these very large boxes. The error bars represent the standard deviation in an ensemble of ten simulations.

We see that $\lambda$ is slightly greater than zero which indicates that there are small departures a the scaling solution (keep in mind that exact linear scaling obviously corresponds to $\lambda=0$). In other words, the wall network is not equilibrating as fast as allowed by causality. Overall we find
\begin{equation}
\lambda_{id,mat}=0.11\pm0.05\,, \label{scalingjmat}
\end{equation}
which we can compare with the value we obtained \cite{AWALL} for 3D $512^3$ matter era simulations of standard domain walls
\begin{equation}
\lambda_{st,mat}=0.04\pm0.02\,. \label{scaling0mat}
\end{equation}
The two values are in fact relatively similar; the larger exponents in the case with junctions should be attributable (at least in part) to the longer time needed for the relaxation to scaling. The fact that as we increase the box size, thus evolving the simulations for a longer dynamic range, $\lambda$ gets closer to zero is a clear indication that the networks are slowly approaching a scaling solution. On the other hand, frustration (which would correspond to $\lambda=1$) is clearly ruled out.

\begin{figure}
\includegraphics[width=3in]{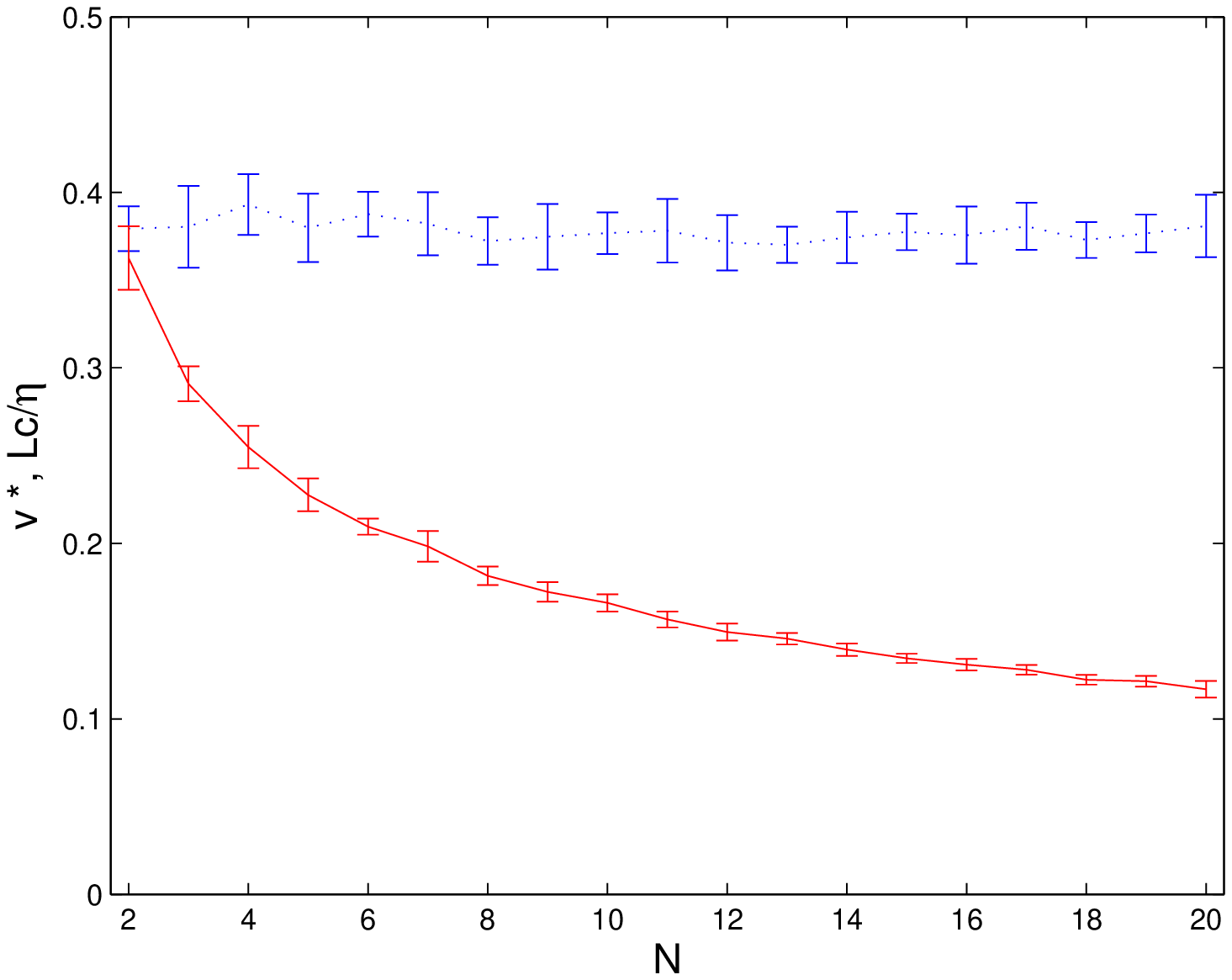}
\includegraphics[width=3in]{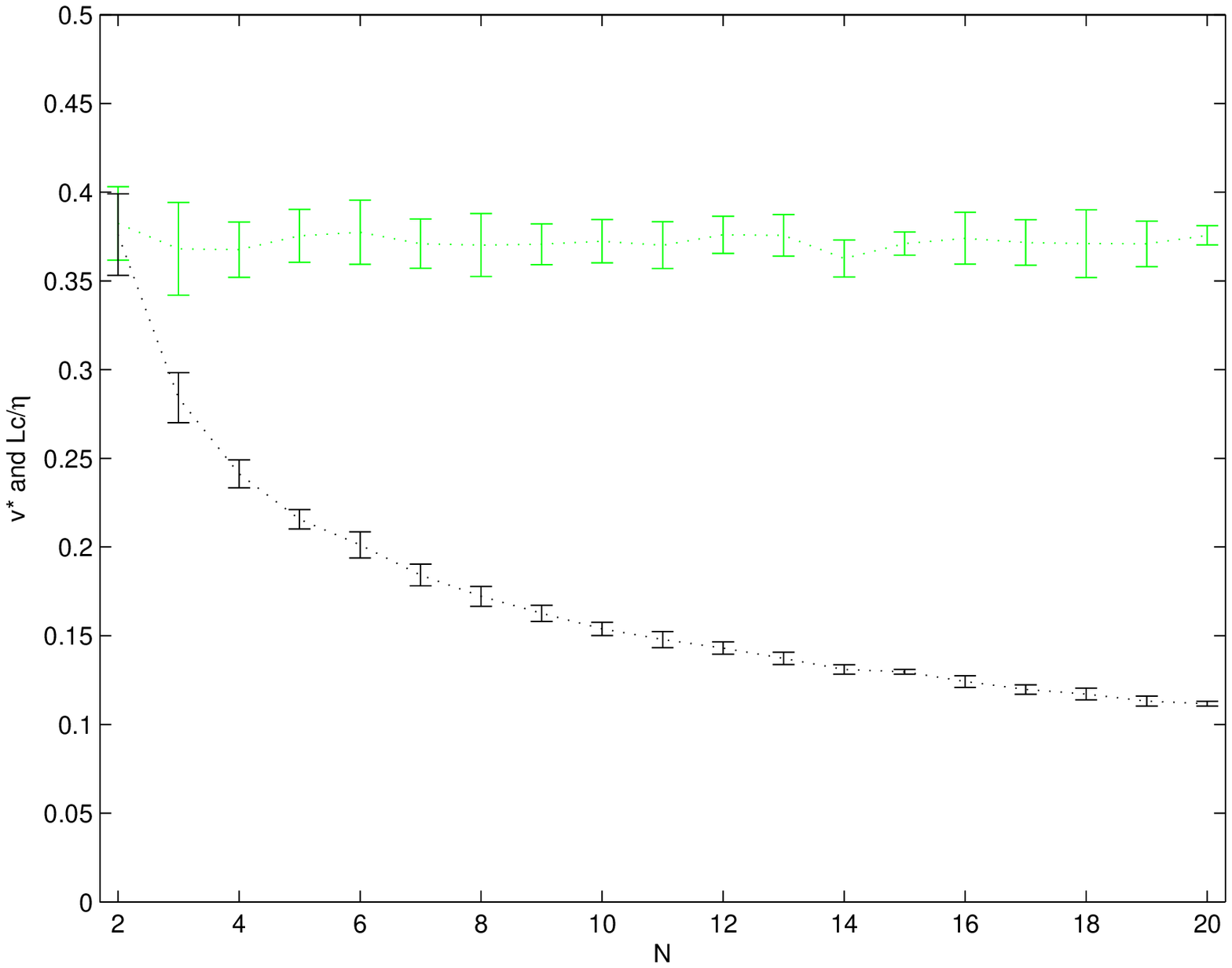}
\caption{\label{scalmat} The asymptotic values of $v_*=\gamma v$ and $L_c/\eta$ (dotted and solid lines respectively) for $256^3$ (top panel) and $512^3$ (bottom panel) matter era simulations of the ideal class of models with $N$ ranging from $2$ to $20$. The error bars represent the standard deviation in an ensemble of 10 simulations.}
\end{figure}

In Fig. \ref{scalmat} we plot the asymptotic values of $v_*=\gamma v$ and $L_c/\eta$ for the above $256^3$ and $512^3$ runs. (Again, the error bars represent the standard deviation in an ensemble of $10$ simulations.) As in the previous plots it is noticeable that the differences between the results obtained for boxes of different sizes are visible but quite small---in general the results are consistent with one another, given the error bars. The way the differences go is also what one expects. Mindful of these differences, we can still say that for many of our purposes the $256^3$ or even $128^3$ boxes (each of which we can usually run in about an hour or less) will produce adequate results.

As we increase $N$ the asymptotic value of $L_c/\eta$ decreases, which is expected as we get closer to the (asymptotic) ideal model. It is also significant that the differences between the successive $N$ results for $L_c/\eta$ become increasingly smaller for large $N$ which is a clear indication that the results obtained for $N=20$ are already reasonably close to the $N \to \infty$ results. On the other hand, we do not find any significant dependence the velocities with $N$. Indeed tipically we find
\begin{equation}
v_{id,mat}=0.36\pm0.02\,, \label{velocjmat}
\end{equation}
where we estimated $v$ as $v_*=v (1-v^2)^{-1/2}$ or equivalently $v=v_* (1+v_*^2)^{-1/2}$. Again we can compare this with the value we obtained for analogous simulations of standard domain walls
\begin{equation}
v_{st,mat}=0.39\pm0.02\,; \label{veloc0mat}
\end{equation}
thus the velocities are slightly smaller in the case of networks with junctions though given the various numerical uncertainties the difference between the two is not very significant. However, the scaling density in the case with junctions is much larger than in the standard case---by a factor of about $3$ for $N=2$, and increasing with the number of fields (we will discuss this dependence later on). A more important comparison can be made with the value we would expect if the network had no energy losses. In this case (recall Eqn. \ref{zerocmat}) we would expect $v\sim0.41$, and the difference means that energy losses are still noteworthy, despite the existence of the junctions. This is ultimately the physical reason why the frustration mechanism can never be realized.

This difference between the behavior of the two averaged quantities is crucial---although changing the number of fields $N$ will change the network's characteristic length (and therefore its density), the fact that the velocities do not change is an indication that in the `local' dynamics of each individual wall until the collpase will not strongly depend on $N$ although the probability that the collapse of a given domain will result in the fusion of adjacent domains will be higher for smaller $N$. Note that we are assuming that the junctions themselves are dynamically unimportant. 

\subsection{Ideal model, other eras}

\begin{figure}
\includegraphics[width=3in]{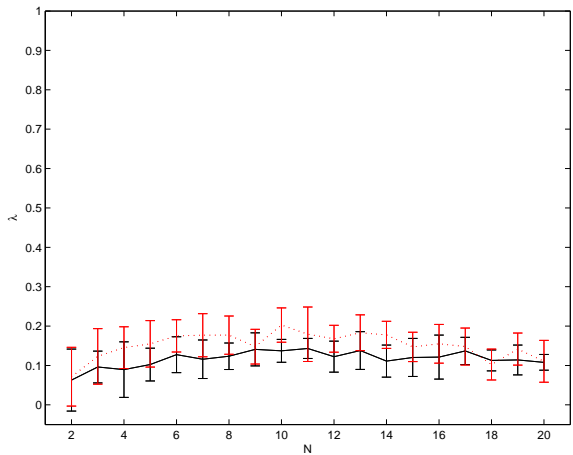}
\caption{\label{exponrad}  The scaling exponents, $\lambda$, for all $N$'s between 2 and 20, for the $128^3$, $256^3$ boxes (dotted and solid lines respectively) in the radiation-dominated era. The error bars represent the standard deviation in an ensemble of 10 simulations. Note that frustration would correspond to $\lambda=1$.}
\end{figure}

\begin{figure}
\includegraphics[width=3in]{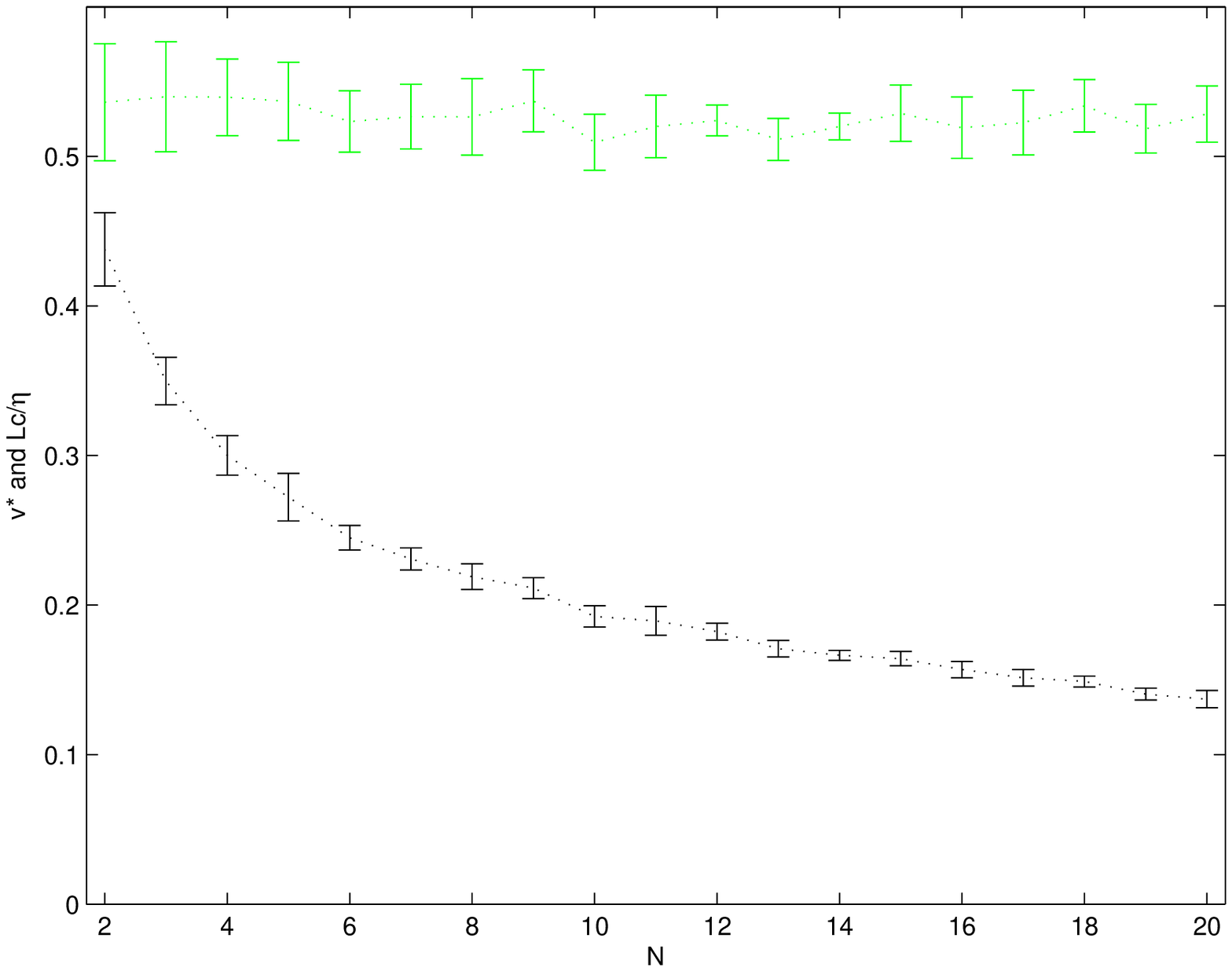}
\includegraphics[width=3in]{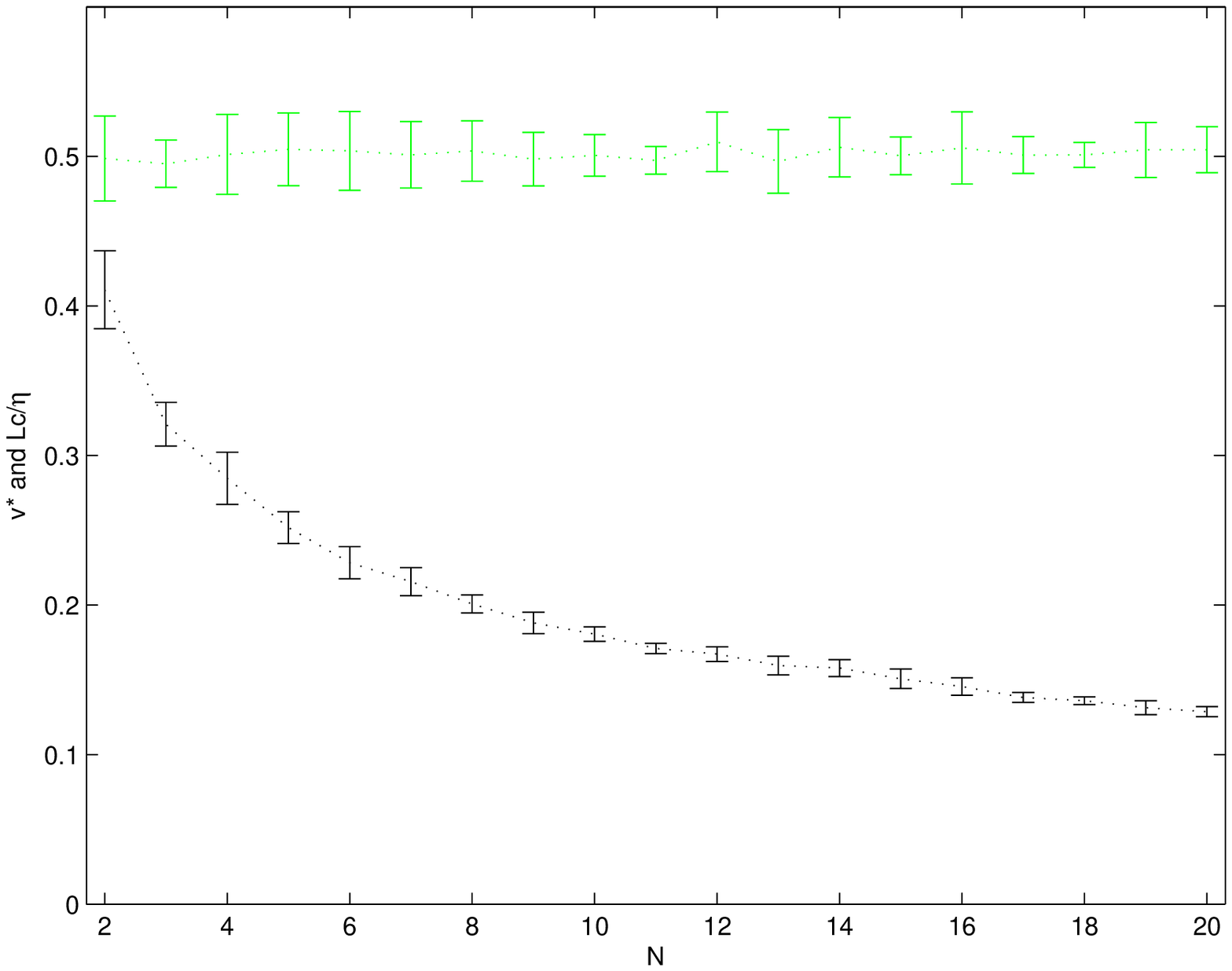}
\caption{\label{scalrad} The asymptotic values of $v_*=\gamma v$ and $L_c/\eta$ for radiation era simulations of the ideal class of models with $N$ ranging from $2$ to $20$ (top and bottom lines respectively). These come from simulations of $128^3$ (top panel) and $256^3$ (bottom panel), and the error bars represent the standard deviation in an ensemble of 10 simulations.}
\end{figure}

Even though simulations of the matter era (that is, with the scale factor evolving as $a\propto t^{2/3}$) are the most relevant from the point of view of cosmological scenarios involving domain walls, it is also interesting to study their evolution in other cosmological epochs. Results of analogous simulations of the ideal model for the radiation era ($a\propto t^{1/2}$) are shown in Figs. \ref{exponrad} and \ref{scalrad}. The scaling exponents and properties are defined and measured as before, and the simulation parameters are also similar, the only difference being that in this case we only have carried out $128^3$ and $256^3$ simulations. 

The results are qualitatively identical to the ones in the matter era, although as expected there are some quantitative differences in the scaling parameters. We now find the scaling exponent
\begin{equation}
\lambda_{id,rad}=0.10\pm0.05\,, \label{scalingjrad}
\end{equation}
which is again comparable to the one obtained for analogous simulations of standard domain walls
\begin{equation}
\lambda_{st,rad}=0.04\pm0.02\,. \label{scaling0rad}
\end{equation}
Likewise, we do not find any significant dependence the velocities with $N$. We now obtain
\begin{equation}
v_{id,rad}=0.45\pm0.03\, \label{velocjrad}
\end{equation}
whereas for standard domain walls in the radiation era
\begin{equation}
v_{st,rad}=0.48\pm0.02\,, \label{veloc0rad}
\end{equation}
and the value we would expect if the network had no energy losses is now $v\sim0.57$ (recall Eqn. \ref{zerocrad}). The only noteworthy difference is that energy losses are comparatively more important in this case---at about the twenty percent level, as opposed to ten percent in the matter era. There is also an enhancement of the scaling density that is similar to the one in the matter era.

\begin{figure}
\includegraphics[width=3in]{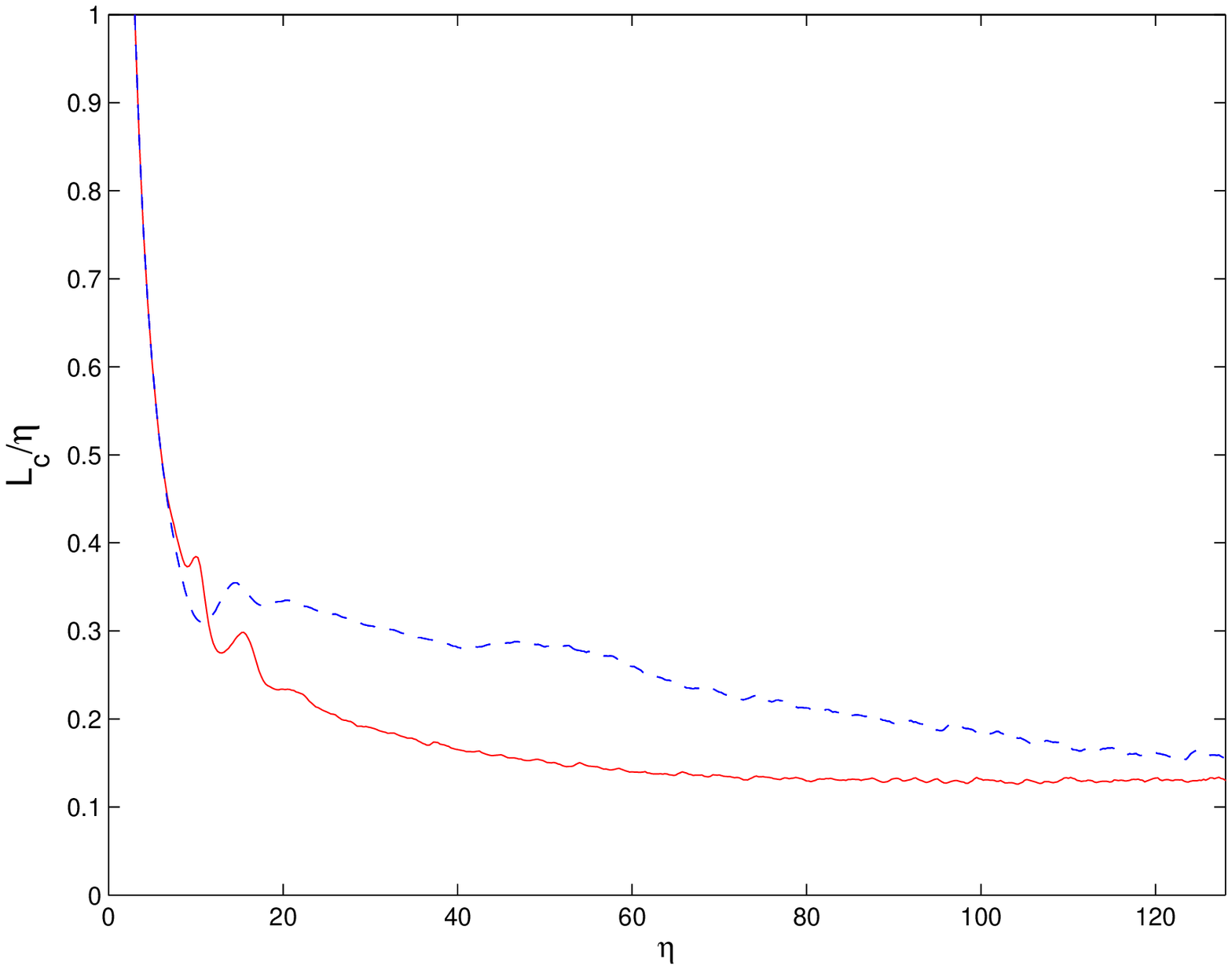}
\includegraphics[width=3in]{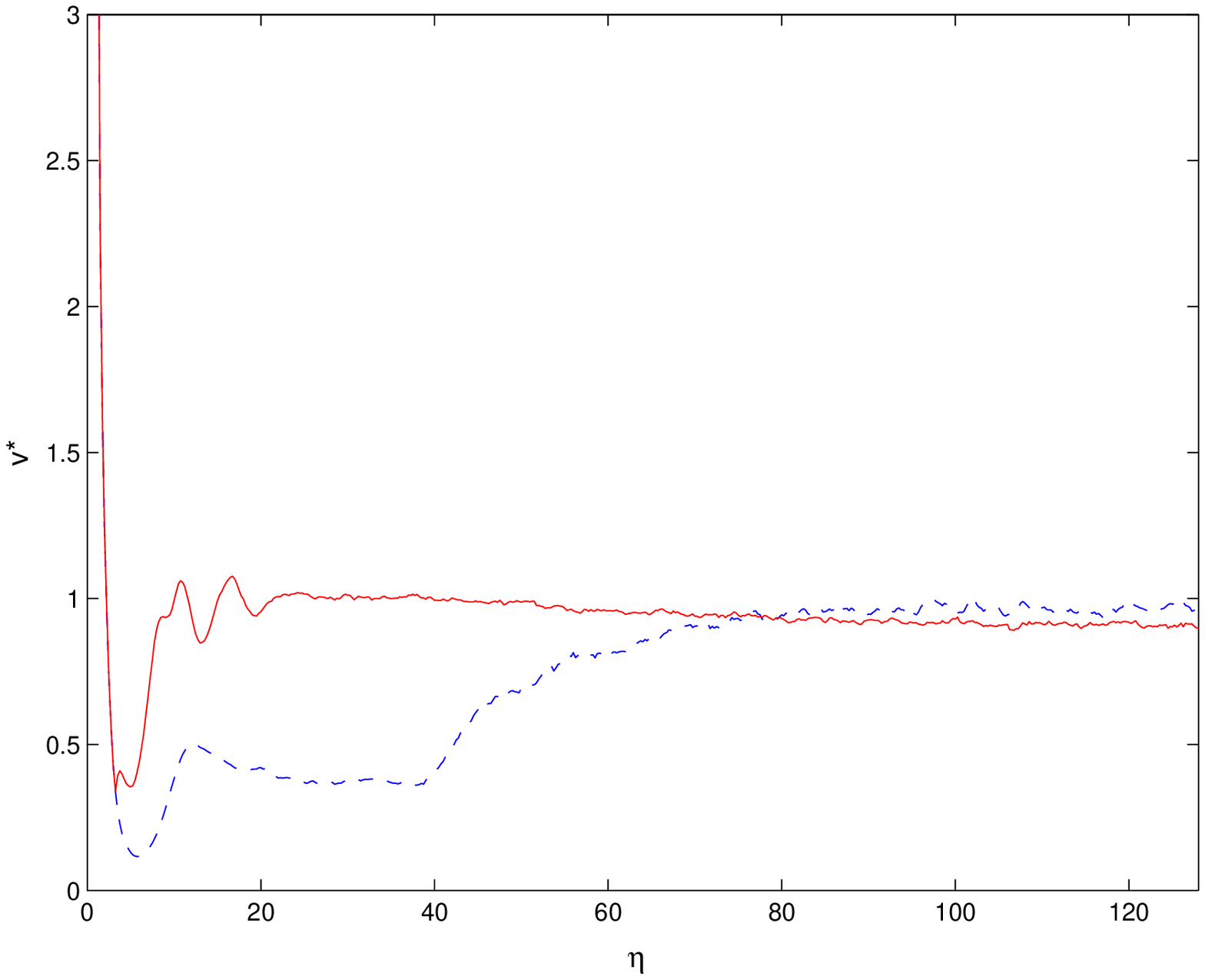}
\caption{\label{scalslw} The asymptotic values of $L_c/\eta$ and $v_*=\gamma v$ for individual $256^2$ simulations of the ideal class of models with $N=3$ in a slow-expansion $a\propto t^{1/5}$ era (solid lines) and in Minkowski spacetime (dashed lines). Note that in both cases the early evolution has a matter-like expansion rate, in order to dissipate the radiation in the box which would otherwise contaminate the measurements.}
\end{figure}

Fig. \ref{scalslw} shows the results of analogous simulations for slower expansion rates, namely a cosmological epoch with $a\propto t^{1/5}$ and Minkowski spacetime (where the expansion is switched off). The former case may be cosmologically relevant in some toy models, for a transient epoch in the very early universe, but in any case it is an important numerical test for our purposes, because our analytic modeling leads us to expect that for $a\propto t^\alpha$ and $\alpha<1/4$ a linear scaling solution can only exist if there are energy losses. 

An accurate numerical evolution of this case is technically difficult, for reasons that are well documented in the analogous case of cosmic strings \cite{Moore}. The slower or non-existing expansion leads to little dissipation, which in turn implies that there is a large thermal bath in the box. In the case of strings, the optimal way to get around this numerical difficulty is to introduce an early period of gradient flow \cite{Moore}. In the present case we adopted the simple solution of evolving the early part of these simulations with a matter-era expansion rate, $a\propto t^{2/3}$, which some testing suggests is accurate enough for our purposes. Naturally, the slower the expansion the more important will be the effect of any radiation present in the box, and therefore the more important it is that this is dissipated in the matter-like evolution phase (which will therefore need to be longer). Our numerical tests suggest that for the $a\propto t^{1/5}$ expansion rate it is sufficient to switch on the matter-like dissipation phase during the initial $2.5\%$ of the simulation, but in the case of the Minkowski simulations this increases to about $30\%$. These values have been used for the simulations shown in Fig. \ref{scalslw} (which correspond the $N=3$ case of the ideal class of models), and the effects of the switch between the early and the late-time evolution are clearly visible in the case of the Minkowski simulation.

Even fairly small $256^2$ simulations are enough to show evidence for an approach to a linear scaling solution in both of these cases. Moreover, velocities become constant, with $v_*\to1$ in the Minkowski spacetime limit, This result therefore confirms the idea that the energy losses are crucially important, even for domain wall networks with junctions, and that these play a fundamental role in the evolution of defect networks. It is possible that the microscopic production of coherent excitations of the underlying quantum field theory \cite{Borsanyi} plays an important role here. This issue is beyond the scope of the present article, but deserves further study.

\subsection{Can the ideal model frustrate?}

Given that we are finding that linear scaling solutions seem to be ubiquitous for the ideal class of models, one might ask if this necessarily so. In other words, could there be some unsuspected pathological feature, hidden in the way the ideal class of models was designed, that would necessarily lead it to linear scaling and artificially prevent it from frustrating? However, it is easy to exclude this possibility. It is well known that domain wall networks should be conformally stretched ($L\propto a$) if their motion is sufficiently damped \cite{AWALL}, for example by friction due to particle scattering in the early stages of their evolution or during an inflationary phase. So a simple way of checking that our ideal class of models can in principle frustrate is to show that it does reproduce this behavior.

\begin{figure}
\includegraphics[width=3in]{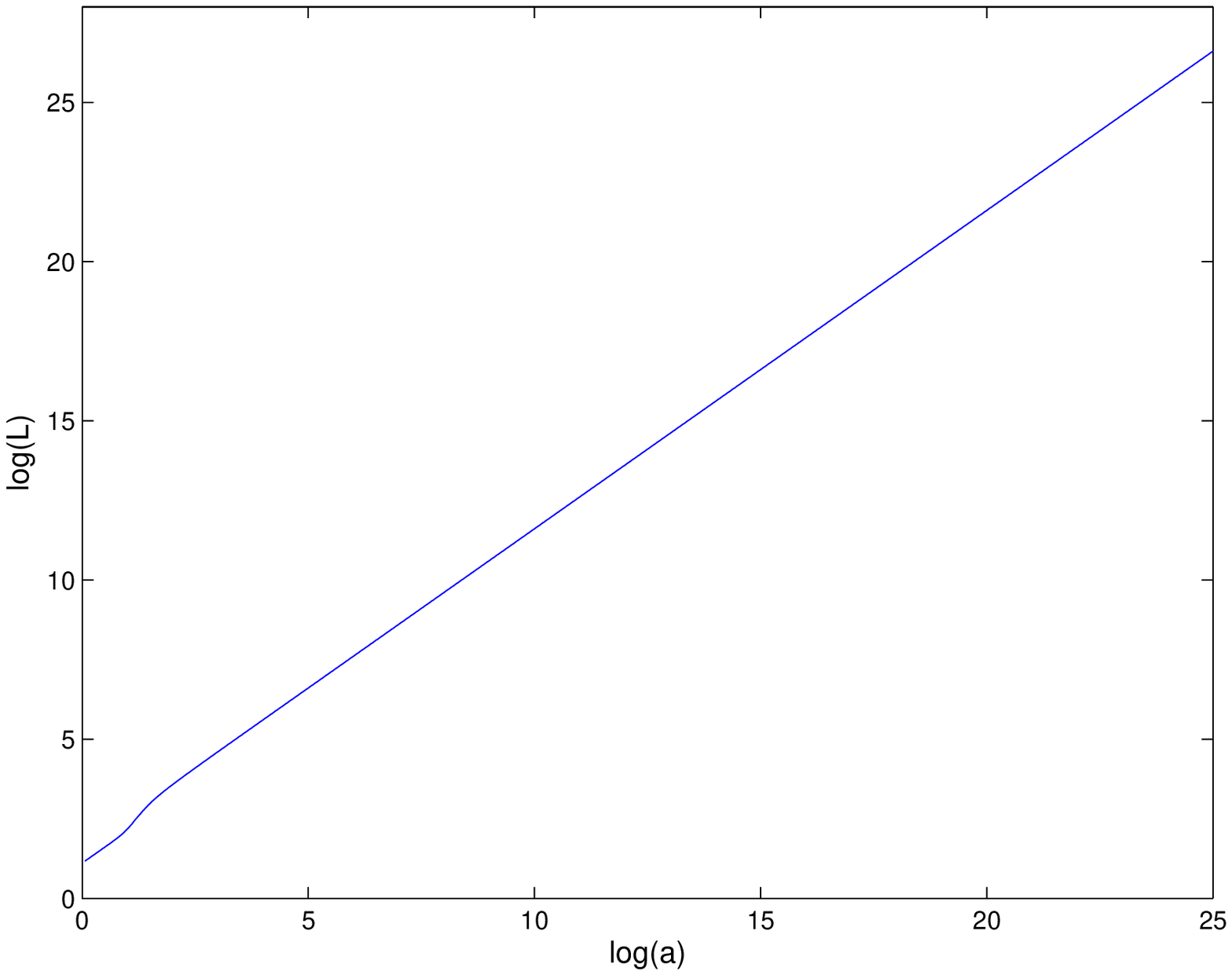}
\includegraphics[width=3in]{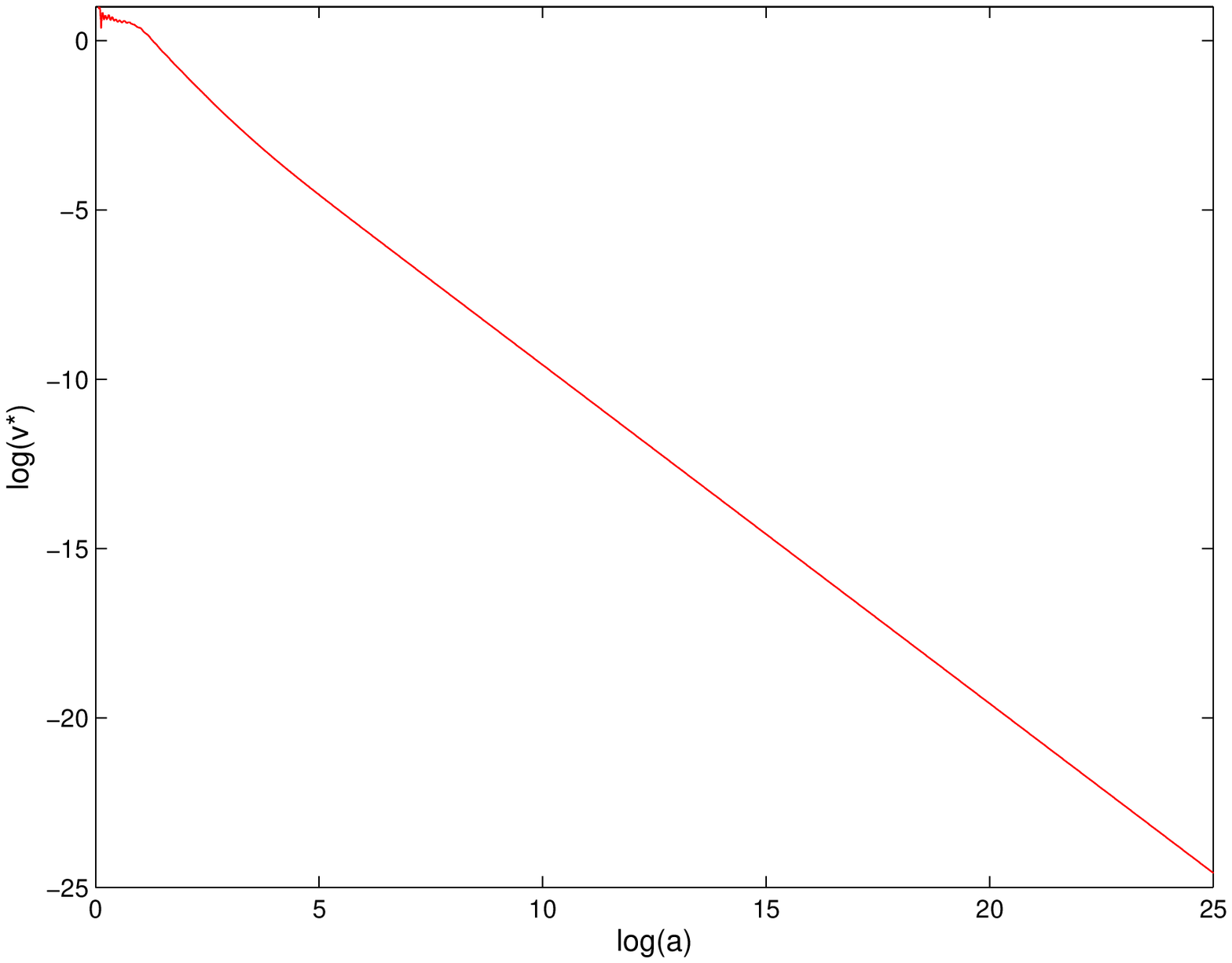}
\caption{\label{exponinf} The evolution of $L_c$ (top panel) and $v_*$ (bottom panel) as a function of the scale factor for a $1024^2$ de Sitter space simulation of the ideal class of models with $N=10$, showing the expected behavior: $L\propto a$, $v\propto a^{-1}$.}
\end{figure}

Performing field theory simulations of this type in de Sitter space involves some numerical subtleties (and, in particular, required some changes in our code) but can nevertheless be done. Fig. \ref{exponinf} shows the results of one such of $1024^2$ simulation, for the $N=10$ case of the ideal class of models. We can immediately see that the ideal class of models does behave exactly as we would expect it to do, namely with $L\propto a$ and $v_*\sim v\propto (HL)^{-1}\propto a^{-1}$. The ideal model is therefore well-behaved and it can in principle frustrate when this is dynamically preferred.

\subsection{The ideal model in other dimensions}

Let us now make a short digression to discuss the dependence of our results on the number of spacetime dimensions. A simple point can illustrate why one might expect the possibility of such a dependence. It is well known \cite{VSH} that if the population probability for a given vacuum is larger than a critical probability $p_c$ (whose value depends on the lattice), that will percolate the lattice; otherwise finite vacuum bags will form. For a 3D cubic lattice $p_c=0.311$, so in the simplest case of a model with two vacua each with $p=0.5$, both will percolate across the lattice. For a 2D square lattice $p_c=0.593$, so in the same situation neither of them percolates. Among other things, this means that standard domain wall networks in 2D and 3D will have very different vacuum topologies. Nevertheless, our previous work \cite{SIMS1,AWALL} shows that this does not seem to affect the dynamical properties of the networks, and in particular its scaling properties.

We now extend this result for the case of domain wall networks with junctions, by providing examples of the fact that the different dimensionality does not seem to affect the frustration conditions. With this aim we will study the behavior of some realizations of the ideal class of domain wall models in two and four spatial dimensions. (In all cases we define domain walls as the topological defects of co-dimension one.) 

\begin{figure}
\includegraphics[width=3in]{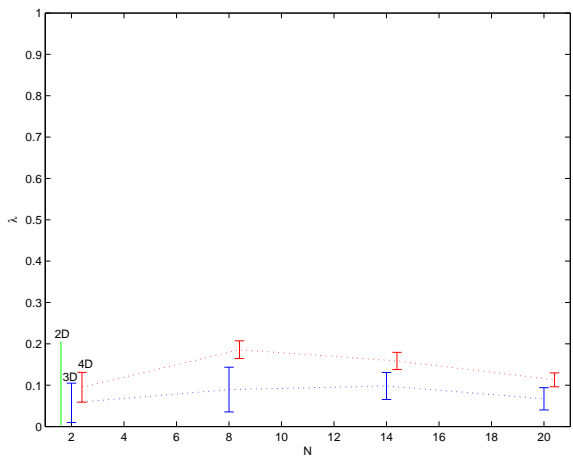}
\caption{\label{dimensions} Comparing the scaling exponents ($\lambda$) for several $N$'s in the matter era for simulation boxes of different dimensions: 2D, 3D and 4D boxes respectively had sizes $8192^2$, $512^3$ and $128^4$. The error bars represent the standard deviation in an ensemble of 10 simulations.}
\end{figure}

Fig. \ref{dimensions} shows a comparison of the the scaling exponents we obtained in our 3D ($512^3$) matter era simulations for $N=2,8,14,20$ (described earlier in the paper) with the corresponding exponents in 4D ($128^4$). A good agreement is found, with the slightly larger exponent and smaller (statistical) error bars being attributable to the smaller dynamic range of the 4D runs. Further evidence for this can be seen by performing 2D runs ($8192^2$), the result of which is also displayed for the case $N=2$.

\subsection{Non-ideal models}

It is also instructive to compare the ideal mode to models where different kinds of junctions can appear. Specifically, we'll consider some realizations of two other classes of models, the Bazeia-Brito-Losano (henceforth referred to as BBL) and Kubotani models \citet{BAZEIA,KUBOTANI} as interesting benchmarks. In previous work these models allowed us to point out the key mechanisms at play, and characterize the differences between models with stable Y-type junctions, models with stable X-type junctions, and models where both types can co-exist. Here we will not describe these models in detail, but only highlight a few of their properties that will be relevant for comparison to the ideal class of model. We refer the reader to the original papers and particularly to \citet{IDEAL2}, for a thorough characterization of these models. We will also revisit some of these properties and extend our previous characterization when discussing the role of junctions in Sect. \ref{hierarchy}.

The 2-field BBL model has minima at the vertices of a square in the plane of the two fields, the orientation of which depends on the model parameters. The model allows for stable $Y$-type or $X$-type junctions depending on the value of the parameter which controls the tension of the walls connecting each pair of vacua. There are two classes of walls which we will denote simply by \textit{edges} and \textit{diagonals}: in the former the wall joins two neighboring minima in field space, while in the latter the wall joins two opposite minima. The fact that this model has parameters that control the ratio of energy between the two types of walls and consequently the type of junctions that are formed makes it very useful to test the energetic and geometric considerations discussed in Sect. \ref{idealmodel}. 

For models with three scalar fields there is an analogous BBL model but an alternative is provided by the (perturbed $O(3)$) Kubotani model. Again there are two branches for the minima, which will be placed either at at the vertices or at the centers of the faces of a cube in the space of the scalar fields. In the former case there are three kinds of walls, which for obvious reasons we can refer to as \textit{edges}, \textit{external diagonals} and \textit{internal diagonals} while in the latter there are two kinds of walls, which we can refer to as \textit{edges} and \textit{axes}. Again the choice of the model parameters determines what type of junctions will be present: in addition to the parameter ranges where only X-junctions or Y-junctions are stable, now there is also a range of parameter where both are stable and coexist. The 3-field BBL and Kubotani models are quite similar, and it is possible to exhibit a correspondence between both potentials, although this is not one-to-one. This implies the Kubotani-model simulations can also be interpreted as BBL model simulations, although the converse need not be true.

\begin{figure}
\includegraphics[width=3in]{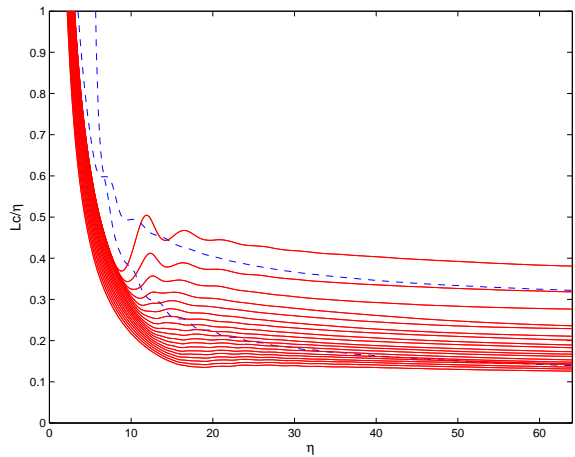} 
\caption{\label{BBLfig} The values of $L_{c}/\eta$ as a function $\eta$ for the ideal class of models (solid lines) with $N$ ranging from $2$ to $20$ (from top to bottom respectively) and for the BBL model with 2 and 3 fields (dashed top and bottom lines respectively). Details of these models can be found in \protect\cite{IDEAL2}. For both BBL models the simulations have been with the model parameter choice $\epsilon r^{2}=-3/10$. All lines correspond to an average of ten $128^{3}$ simulations.}
\end{figure}

Our present intention is to compare the ideal model with these other models from the point of view of producing frustrated networks. We have performed series of  $128^{3}$ matter-era simulations of the BBL model in the case where only $X$-type junctions are allowed. In both the 2-field and the 3-field case for the parameter choice $\epsilon r^{2}=-3/10$ will lead to this scenario---again, refer to \cite{IDEAL2} for a discussion of this parameter.

In Fig. \ref{BBLfig} we compare the values of $L_{c}/\eta$ as a function $\eta$ for the ideal class of models, with $N$ ranging form $2$ to $20$, with the aforementioned BBL models with stable $X$-type junctions. As expected, we find that the 3-field BBL model gives a better result than the 2-field one, but there is no significant improvements with respect to the ideal class of models. The no frustration conjecture is evident for all models. 


\section{\label{smod}Further modeling}

In this section we will provide an analysis of some of the above results based on the simple analytic model which we introduced in Sect. \ref{vosmodel}. The model was derived and extensively described in \cite{AWALL}, and is analogous to an earlier model for cosmic strings \cite{QUANT}. There it was also shown that it provides a good description of the evolution of the large-scale properties of standard domain wall networks. We also discuss the behavior of this type of networks in the ideal limit $N\to\infty$.

\subsection{The low-$N$ case}

For standard domain walls a reasonable fit to the numerical simulations, both in the radiation and in the matter eras, is obtained for the following choice of model parameters \cite{AWALL}
\begin{equation}
k_0 \sim 1 \,,\quad c_0\sim k_0/2\,. \label{oldfit}
\end{equation}
Note that since this is a fairly simple phenomenological model we won't include any error bars in the model parameters; these would typically be at the ten to twenty percent level.

Let us start by concentrating on the simulations of the case $N=2$. Since the late-time evolution is found to be close to the linear scaling regime, we can approximately use the scaling relations given by Eqns. \ref{defscalingg}--\ref{defscalingv}, and we then find that the analytic model still provides a good fit to the data, although the phenomenological parameters are now
\begin{equation}
k_2\sim0.7\,,\quad c_{2}\sim  k_2/2\, \label{newfit}\,.
\end{equation}
This holds for both the matter and radiation eras, although the fit is better in the former case than in the latter. This bias towards the matter era simulations in the choice of fit parameters is deliberate, since we have done larger simulations in the matter era, and the faster expansion rate (relative to that the radiation era) also means that these simulations should be less affected by the thermal bath in the box.

As was to be expected, we find parameter values that are significantly smaller than in the case of standard domain walls. The presence of junctions means that the walls tend to be straighter, and hence $k$ (which is to some extent has the phenomenological role of a curvature parameter) must be smaller than in the standard case. Nevertheless, note that it is still quite close to unity--- our earlier work \cite{IDEAL1,IDEAL2} shows that it would need to be several orders of magnitude smaller for frustration to be viable. 

Now let us turn our attention to the behavior of the scaling properties as we increase the number of fields. We have already pointed out when presenting the simulation results that the fact that the scaling velocities are independent of the number of fields is not really surprising. The key features of the ideal class of models are that all the minima of the potential are equally spaced and all the domain walls have the same tension (a model with $N$ fields will have $N+1$ minima). These in turn necessarily imply \cite{IDEAL1} that the junctions will always be of $Y$ type, regardless of the number of fields. Naively, on therefore expects that in a statistical sense all walls of a given network (in a given cosmological epoch and with a given number of scalar fields) probe a similar region of the potential, and will have the same basic properties, such as the characteristic scale and root-mean squared velocity. In these circumstances, recalling our discussion in Sect. \ref{basicdef} and provided we can assume that the junctions themselves do not have any crucial dynamical role, then the local dynamics of a single domain wall will effectively be roughly the same regardless of the number of fields. We will return to the issue of the dynamical role of the junctions in the following section.

As an aside, recalling that the equation of state of a domain wall gas is given by
\begin{equation}
w \equiv \frac{p}{\rho} = -\frac{2}{3} + v^2\,,
\end{equation}
our results on the network velocities allow us to predict the following equations of state for scaling domain wall networks with junctions in the radiation and matter eras
\begin{equation}
w_{rad}\sim-\frac{7}{15}\,,\quad w_{mat}\sim-\frac{13}{24}\,,\label{epochvs}
\end{equation}
so the correct value is closer to $w\sim-1/2$ than to the naively expected $w=-2/3$. We emphasize that for the (ideal) class of models we are considering, these will hold for any number of fields. It's also clear that these are in conflict with the existing observational bounds.

\subsection{$N$ dependence and the ideal limit}

Changing the number of fields $N$ will change the network's characteristic length, and hence its density. Looking at the plots for the scaling density parameter $\epsilon=L/t$ in the matter and radiation eras, it is strikingly clear that
\begin{equation}
\epsilon_{low} (N)\propto \frac{1}{\sqrt{N}}\, \label{scalingfields1}
\end{equation}
provides a good fit to the data. If this fit were to hold for any number of fields then it would obviously imply that $\epsilon_N\to0$ as $N\to\infty$. In other words, even for a fixed defect tension one could freely tune the energy density of the network at a given subsequent epoch to take any value we wanted, simply by choosing the appropriate number of fields---although this would be a clear form of fine-tuning. However, it is clear that Eqn. \ref{scalingfields1} is not valid for all values of $N$, and specifically in the limit $N\to\infty$. In fact in the large $N$ limit we expect a behavior of the form
\begin{equation}
\epsilon_{high} (N)=A+\frac{B}{N}\,, \label{scalingfields2}
\end{equation}
corresponding to a constant asymptotic density. The reason is clear. On one hand, $L/(vt)$ has to be of order unity and we have seen that the characteristic velocity of the domain walls, $v$, is rather large and appears to be independent of $N$.  On the other hand the probability that the collapse of a given domain results in the fusion of adjacent domains is proportional to $1/N$, for large $N$. The fusion of adjacent domains will change the domain wall density with respect to that of the $N\to\infty$ limit by a factor proportional to the probability of such events, if $N$ is large, which leads to  Eqn. \ref{scalingfields1} in this limit.

However, this $N$-dependence overestimates $\epsilon$ (thus underestimating the density) for small numbers of fields, but an adequate fit can be obtained for the larger values we simulated. Specifically, in the range $10\le N \le 20$ we find
\begin{equation}
B_m\sim1.0\,,\quad B_r\sim1.2\,, \label{coeffb}
\end{equation}
respectively in the matter and radiation eras, while in both cases
\begin{equation}
A\sim0.06\,, \label{coeffa}
\end{equation}
which would therefore characterize the asymptotic scaling law. 

Regardless of the detailed dependence on the number of fields, for sufficiently large energy densities the wall network will at some point become the dominant energy component, and the subsequent evolution of the universe will be dramatically altered---see \cite{AWALL} for a brief discussion of this scenario. And even before this stage is reached, a sub-dominant but significant contribution to the total energy density might have consequences that will conflict with existing observations---we will return to this point in Sect. \ref{scon}. Before that, however, we will consider what changes (if any) are to be expected in the above results if the junctions are dynamically important.


\section{\label{sjun}The role of junctions}

The types and properties of junctions that can form will depend both on energy considerations and on the topology of the potential's minima in field space. Since at least $n$ real scalar fields are required to produce a defect of co-dimension $n$, it follows that a junction of the respective type will correspond to an $n$-dimensional field space configuration, connecting at least $(n+1)$ minima. Note that one can have configurations where defects intersect but there are no physical junctions (a trivial example is the case of decoupled fields). One can then discuss examples of particular models, as we will do below. We will then briefly address the issue of the role of the junctions themselves in the dynamics of the defect networks.

\subsection{\label{hierarchy}Examples of junction hierarchies}

In what follows we will discuss the possible junction hierarchies in three classes of models we have been considering: in addition to the ideal class we have been discussing for most of this article, we will again use the BBL and Kubotani models \cite{BAZEIA,KUBOTANI} as interesting benchmarks. The present brief discussion can be seen as an extension of that in \cite{IDEAL2} (we leave a more detailed analysis, particularly of energetic arguments, for future work). Here as in that article, the point we wish to emphasize is that despite very simple rules for the formation of junctions, a combination of topological, geometric and energetic arguments can lead to quite different behaviors, depending on the details of the particular model. As one might expect, the ideal class of models will turn out to have the simplest and physically clearer phenomenology. Notice that for the sake of clarity the discussion will focus on the case of 3 spatial dimensions---however, it's simple enough to repeat for other dimensions.

In the $2$-field BBL model there are always 4 minima, placed at the vertices of a square. Domain walls form, and intersect at string-type junctions. Depending on the model parameters, energetic configurations will imply that either X-type or Y-type wall junctions will be stable (that is, with either 4 or 3 walls meeting at each junction)---see \cite{IDEAL2} for a discussion of both cases. There is no situation in which both are simultaneously stable. Notice that in the former case there is only one type of string that can be formed, while in the latter there are 4 types, although they all have the same energy. Also notice that in this model there are physically no monopole-type junctions---which is obvious given that there are only two scalar fields.

\begin{figure}
\includegraphics[width=3in]{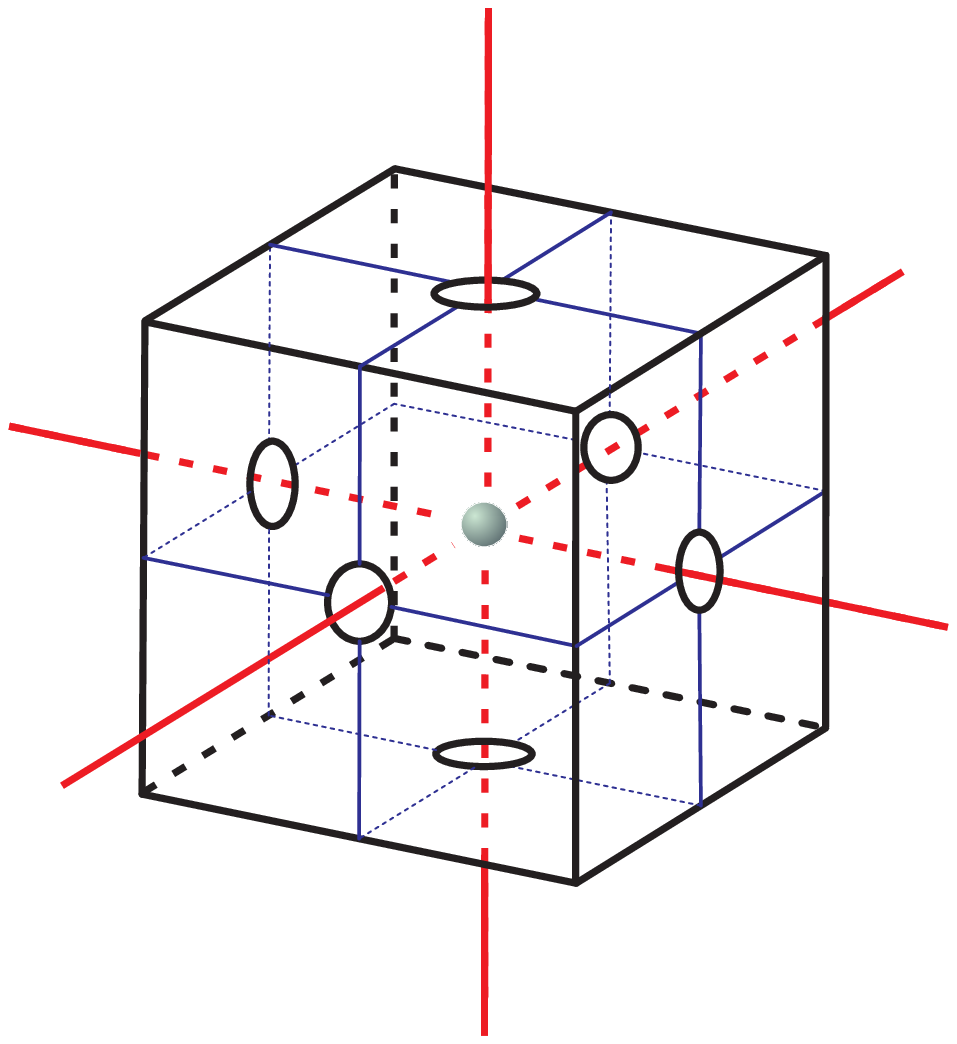}
\includegraphics[width=3in]{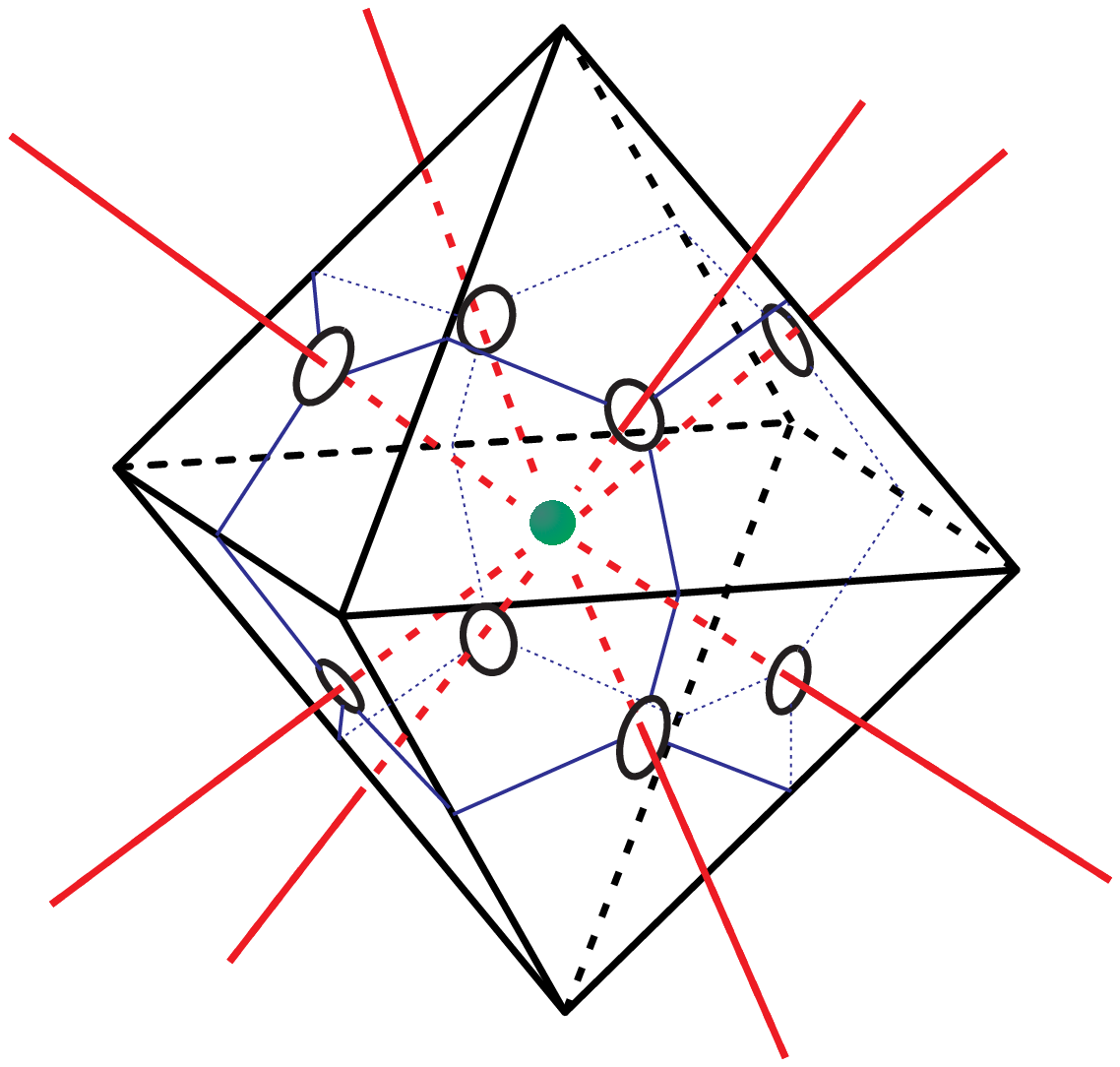}
\caption{\label{hierarchykub} Two possible junction hierarchies for the Kubotani model in three spatial dimensions. In the so-called positive branch (top panel) 4 walls meet at a string-type junctions and 6 strings meet at a monopole-type junction. In the negative branch (bottom panel) 3 walls can meet at a string-type junctions and 8 strings can meet at a monopole-type junction. In the latter case other types of junctions are also aallowed---refer to the main text for a discussion.}
\end{figure}

In the positive branch of the Kubotani model (which corresponds to the intermediate range of the BBL $3$-field model), there are 3 scalar fields and 8 minima which occupy the vertices of a cube. Walls form string-type junctions which, by energetic arguments are always of X-type (the Y-types being unstable in this regime), so they form along the edges of the cube. Hence 4 walls meet at a string-type junctions, but now the hierarchy continues one further step, and 6 strings meet at a monopole-type junction. Notice that there is only one type of (stable) monopole, where 12 walls effectively meet (and \textit{not} 24---each wall is effectively contributing to two strings). This is illustrated in the top panel of Fig. \ref{hierarchykub}.

The negative branch of the Kubotani model (which is also the negative range of the BBL $3$-field model) is slightly more subtle. Here there are 6 minima at the vertices of an octahedron. Energetic arguments show that we now have walls with Y-type junctions forming along the octahedron's edges, so 3 walls can meet at a string-type junctions and 8 strings can meet at a monopole-type junction. This is illustrated in the bottom panel of Fig. \ref{hierarchykub}. However, in this case there are also other types of junctions allowed, both for strings and monopoles. It turns out that energetic considerations also allow for walls to form X-type junctions, as well as Y-type junctions involving the axis sector---both of these have in common the fact that they are configurations in which one of the fields vanishes. From this it follows that there are also monopole-type junctions formed at the intersection of 5 strings (these have 4 strings formed from edge sector Y-type junctions and one formed from an axis sector X-type junction) or at the intersection of 4 strings (this involves 2 strings formed from edge sector Y-type junctions and 2 strings formed from axis sector Y-type junctions). Pictorially these monopoles correspond to topological configurations involving one half or one quarter of the octahedron respectively (or equivalently, square-based and triangular-based pyramids), while the 8-string monopole corresponds to the full octahedron. The relation between the energies of the different monopole-type configurations is not immediately clear---in particular, it is conceivable that some are unstable and will tend to decay.

The positive range of the BBL $3$-field model (which has no correspondence in the Kubotani model) is even more complicated. It has 8 minima at the vertices of a cube (just like the intermediate range) but now energetic arguments imply that the walls meet at Y-type junctions (the X-type ones are unstable in this parameter range). Hence string-type junctions form at the intersection of 3 walls, but one can have three physically distinct classes of junctions: they can involve 3 walls from external diagonals, or 2 walls from edge sectors and one from an external diagonal, or one wall from each of the sectors (edge, external diagonal and internal diagonal). We can refer to each of these strings as Type I, II and III respectively, according to the number of different wall types involved. Assuming that all three classes are allowed, monopole-type junctions can form at the intersection of 4 strings, and again there will be 3 possible types of monopoles. Interestingly, they all must necessarily involve a combination of different types of the wall junctions: one can have monopoles at the intersection of three Type II and one Type I string, or two Type II and two Type III strings, or finally one Type I, one Type II and two Type III strings. In any case, 6 walls effectively meet at a monopole, as opposed to the 12 of the intermediate range discussed above. Of these 6, there will be at least 2 walls from edge sectors and 2 from external diagonals. The remaining 2 will be of different types: in the first case we discussed we have one edge and one external diagonal, in the second case we have one edge and one internal diagonal, and in the third case one external diagonal and one internal diagonal.  It would be interesting to study the relative energies of all these configurations in more detail, and hence their stability, both analytically and numerically. We expect that the three monopole cases we listed abovr are ordered by increasing energy.

\begin{figure}
\includegraphics[width=3in]{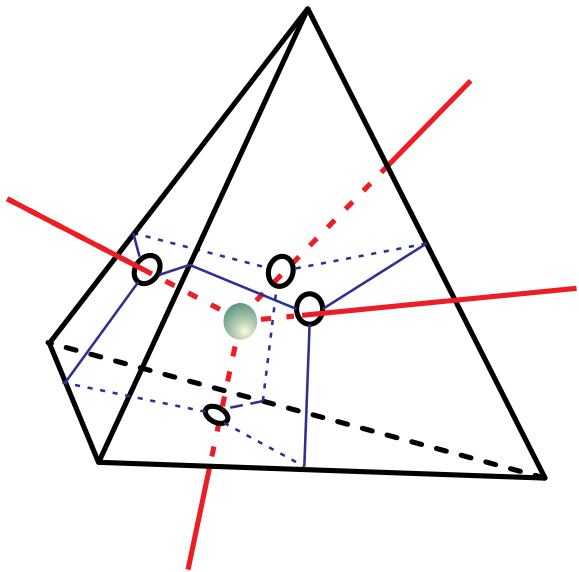}
\caption{\label{hierarchyideal} The junction hierarchy in the ideal class of models, in three spatial dimensions. String-type junctions form at the intersection of 3 walls, and monopole-type junctions form at the intersection of 4 strings.}
\end{figure}

Finally, in the ideal model energetic arguments imply that walls always form Y-type junctions, and with $p$ scalar fields there will be $(p+1)$ minima evenly spaced on the surface of a $p$-dimensional sphere, from which we will produce defects of co-dimension up to $p$. Hence in three spatial dimensions string-type junctions form at the intersection of 3 walls (provided there are at least 2 fields) and monopole-type junctions form at the intersection of 4 strings (provided there are at least 3 fields). An illustration of this configuration is in Fig. \ref{hierarchyideal}. In models with additional dimensions there can be a much deeper hierarchy: the rule is that co-dimension $(p+1)$ defects will form at the intersection of $(p+2)$ defects of co-dimension $p$ (for $p>0$).

\subsection{Junctions and network dynamics}

From the above discussion of the ideal model it's noteworthy that the case $N=2$ of the ideal model is special in that it contains no true monopoles, while those with $N>2$ do have them. (Trivially, the case $N=1$ has neither monopoles nor strings, only plain domain walls.) It is therefore interesting to compare this case with say $N=3$: if the differences are larger than expected from the discussion in the previous sections, this could be an indication of an effect of the monopoles on the dynamics of the network. In order to have a closer look at this, we have carried out $1024^3$ matter-era numerical simulations of the $N=2$ and $N=3$ cases.

\begin{figure}
\includegraphics[width=3in]{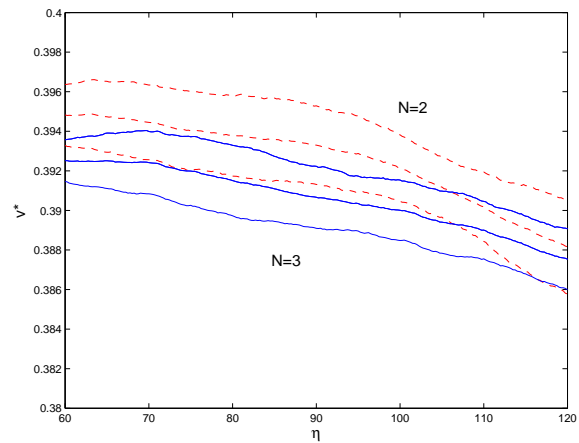}
\caption{\label{monopoles} The averaged velocities (with statistical error bars) of $1024^3$ matter simulations of the ideal class of models with $N=2$ (dashed lines) and $N=3$ (solid lines). The velocities are plotted as a function of the simulation time, and in each case the averages are over 10 simulations.}
\end{figure}

Fig. \ref{monopoles} shows the evolution of the velocities in the two cases (averaged for series of 10 runs of each type). The fact that the velocities are statistically the same (notice the very narrow range of the $y$ axis) naively suggests that the dynamical effect of the monopoles is negligible. Nevertheless, it is of interest to consider what changes would be expected if the junctions are dynamically important. It's clear that this will increase the effective equation of state of the defect network and therefore it will not be any help for dark energy and frustration scenarios, but it might have other interesting consequences.

Very simple models can be used to study this issue---even 2-field models and 2D simulations are sufficient to phenomenologically study the effect of the junctions. The idea is to consider walls which have fixed tension and thickness, but also a junction energy that is a function of a single tunable parameter. One way to implement this would be a perturbed version of the $N=2$ ideal model. An alternative is a to take a perturbed $Z_3$ model, namely
\be
L= \frac{1}{2}\lambda ( \varphi^3 - 1)({\bar\varphi}^3 - 1) + \theta (\varphi^2-1)^2
\ee
where $\theta$ is a positive parameter which controls the junction energy. However, notice that in order to keep a fixed thickness we must modify this to
\be
L= \frac{1}{2}\lambda*f(\theta) ( \varphi^3 - 1)({\bar\varphi}^3 - 1) + \theta (\varphi^2-1)^2
\ee
where $f(\theta)$ is a function to be determined numerically. One can then study the scaling properties and/or the effective equation of state as a function of the parameter $\theta$. This will be left for future work.


\section{\label{scon}Discussion and conclusions}

In this paper we have built upon earlier work \cite{IDEAL1,IDEAL2,THIRD} and presented compelling evidence that domain wall networks cannot be the dark energy. Note that in order to be able to rule out the domain wall scenario for dark energy a very large and rich class of models had to be analyzed in detail. This led us to develop a model best suited for frustration (the 'ideal' model). We have shown that even this model fails to produce a frustrated domain wall network. Current observational constraints using cosmic microwave background and supernova data already strongly disfavor $w=-2/3$ as the equation of state of a single dark energy component \cite{WMAP5}. However, we should bear in mind that these observational results are dependent on strong priors. Even if we take them for granted and accept that domain walls \textit{alone} cannot be the dark energy, die-hards might argue that they could still make a significant partial contribution. Our results, however, exclude even that rather more contrived possibility. 

More generally, we have also provided a quantitative characterization of the cosmological evolution of the ideal class of domain wall networks with junctions. These have the interesting feature that their characteristic velocity appears to be independent of the number of scalar fields $N$. We have also shown how a simple analytic toy model can provide an adequate description of these models and discussed how the scaling density depends on the number of fields, both for the numerically simpler case of a small number of fields and for the physically more interesting case $N\to\infty$. We have also discussed the possible hierarchies of junctions in the ideal class of models, contrasted it with what is found in other models and  briefly addressed the role of the junctions in the network dynamics.

We are now in a position re-examine the possible cosmological roles of domain wall networks and the corresponding bounds on relevant model parameters in the light of our findings.  The Zel'dovich bound \cite{ZEL} tell us that the energy scale $T_{SB}$ of the symmetry breaking phase transition that produced the domain walls has to be smaller than about $1 \ {\rm MeV}$. However, the classic derivation of the Zel'dovich bound implicitly assumes the linear scaling solution with roughly one defect per Hubble volume. As we saw above this is not in general a valid assumption, and in particular it is quite wrong if the domain walls are to have any chance of providing an interesting contribution to the dark energy. 

We have shown that $L \lsim 10 \, {\rm kpc}$ at the present time and consequently $LH \lsim 10^{-6}$ if the domain walls are to provide a significant contribution to the dark energy. Of course since the characteristic scale of the domain walls has to be smaller than the Hubble scale by a factor of at least $10^{-6}$ then the domain wall tension also has to be smaller by a similar factor (with respect to the Zel'dovich bound) in order that domain wall density does exceed the critical density. This translates into a bound on the scale of the symmetry breaking phase transition that produced the domain walls of 
\begin{equation} 
T_{SB} \lsim 10 \, {\rm keV}\,. 
\end{equation}

Finally, let us point out that our massively parallel domain walls code, which has been gradually optimized and generalized to be able to deal efficiently with multi-field scenarios, can have other uses beyond the study of domain wall networks (with or without junctions). With suitable changes it can be used to phenomenologically study cosmic (super)strings, which are generically expected to have junctions, and also to look into a number of string landscape scenarios, where the key feature is the presence of a large number of (usually) coupled fields. Some of these topics will be the subject of future publications.

\section*{Acknowledgments}
C.M. is funded by a Ciencia2007 research contract. J.M. and R.M. are supported by the Brazilian government (through CAPES-BRASIL), specifically through grants BEX-1970/02-0 and BEX-1090/05-4.

The numerical simulations were performed on COSMOS, the Altix4700 owned by the UK Computational Cosmology Consortium, supported by SGI, Intel, HEFCE and PPARC. The support of the COSMOS parallel programmer, Victor Travieso, with the code optimization and data visualization is greatly appreciated.


\bibliography{prd}

\end{document}